\documentclass[12pt,preprint]{aastex}
\usepackage[numberedappendix]{emulateapj5}
\newcommand{\lsim}{\stackrel{\scriptscriptstyle <}{\scriptstyle {}_\sim}}
\newcommand{\gsim}{\stackrel{\scriptscriptstyle >}{\scriptstyle {}_\sim}}
\newcommand{\et}{\mbox{et~al.\ }}
\newcommand{\eg}{\mbox{e.g.,}\ }
\newcommand{\ie}{\mbox{i.e.,}\ }

\newcommand{\kms}{\ifmmode {\rm km\,s}^{-1} \else km\,s$^{-1}$\fi}
\newcommand{\kmsMpc}{\ifmmode {\rm km\,s}^{-1} {\rm Mpc}^{-1} \else km\,s$^{-1}$\,Mpc$^{-1}$\fi}
\newcommand{\ergs}{\ifmmode {\rm ergs\,s}^{-1} \else ergs\,s$^{-1}$\fi}
\newcommand{\lya}{\ifmmode {\rm Ly\,}\alpha \, \else Ly\,$\alpha$\,\fi}
\newcommand{\nv}{N\,{\sc v}}

\newcommand{\feii}{Fe\,{\sc ii}} 
 
\newcommand{\civ}{C\,{\sc iv}} 
\newcommand{\cii}{C\,{\sc ii}}

\newcommand{\siiv}{Si\,{\sc iv}}
\newcommand{\siii}{Si\,{\sc iii}}
\newcommand{\alii}{Al\,{\sc ii}}
\newcommand{\aliii}{Al\,{\sc iii}}

\newcommand{\mgii}{Mg\,{\sc ii}}
\newcommand{\mgi}{Mg\,{\sc i}}
\newcommand{\lam}{$\lambda$}

\newcommand{\mdotratio}{$\dot{M}_{\rm BH}/\dot{M}_{\rm Edd}$}

\newcommand{\hst}{{\it HST}}

\newcommand{\half}{\mbox{$\frac{1}{2}$}}
\newcommand{\degs}{\mbox{$^{o}$}}

\input epsf

\begin{document}
\submitted{Accepted by ApJ, August 20, 2003}
\journalinfo{To appear in The Astrophysical Journal, 599, December 2003}

\title{Occurrence and Global Properties of Narrow C\,{\small IV} 
\mbox{$\lambda$}1549\AA\ Absorption Lines in Moderate-Redshift 
Quasars}

\shorttitle{Quasar Narrow C\,{\small IV} Absorption Lines}
\shortauthors{Vestergaard}

\author{M.\ Vestergaard}
\affil{Department of Astronomy, The Ohio State University,
	140 West 18th Avenue, \\ Columbus, OH 43210-1173.
	Email: vester@astronomy.ohio-state.edu
}

\begin{abstract}
A statistical study is presented of (a) the frequency of narrow \civ{} 
\lam 1549 absorption lines in \mbox{$1.5\,\lsim\,z\,\lsim\,3.6$} radio-quiet 
and radio-loud quasars, and of (b) the UV and radio properties of the 
absorbed quasars. The quasar sample is unbiased with respect to 
absorption properties and the radio-quiet and radio-loud subsamples
are well matched in redshift and luminosity.
A similarly high incidence ($\gsim$50\%) of narrow \civ{} absorbers
is detected for the radio-quiet and radio-loud quasars, and a constant 
$\sim$25\% of all the quasars, irrespective of radio type 
display {\it associated}  \civ{} absorbers stronger than $EW_{\rm rest} \geq$ 0.5\AA{}.
Both radio-quiet and radio-loud quasars with narrow absorption lines have
systematically redder continua, especially strongly absorbed objects.
There is evidence of inclination dependent dust reddening and absorption
for the radio quasars.
An additional key result is that the most strongly absorbed radio quasars 
have the largest radio source extent. This result is in stark contrast to a 
recent study of the low-frequency selected
Molonglo survey in which a connection between the strength of the narrow
absorbers and the (young) age of the radio source has been proposed.
The possible origin of these discrepant results is discussed and may
be related to the higher source luminosity for the quasars studied here.
\end{abstract}


\section{Introduction and Motivation \label{intro}}

Quasar absorption lines fall into two gross categories based on their
line widths, namely narrow and broad. The narrow absorption lines (NALs) have 
line widths less than a few hundred \kms{} and tend to have relatively 
sharp profiles.  The broad absorption lines (BALs) are much more dramatic, 
displaying deep, broad, and smooth absorption troughs extending blueward 
of the emission line profile of the absorbing species; some BAL troughs 
are even `detached' from the line emission profile. In the case of BALs
there is no doubt that the absorbing gas is associated with the quasars
as it is outflowing at sub-relativistic speeds: outflow velocities of 
order $-$20\,000\kms{} to $-$30\,000\,\kms{} are not uncommon, and
velocities reaching $-$60\,000\,\kms{} have been detected (Jannuzi \et 1996;
Hamann \et 1997).
The NALs may be due to intervening galaxies along our line of sight
to the quasar, or due to gas somehow associated with the quasar system.
Gas in the quasar environment, such as a cluster, has been suggested
early on to explain some of the absorbers (\eg Weymann \et 1979; Briggs,
Turnshek, \& Wolfe 1984).  However, gas clearly intrinsic to the quasar is 
also established to produce NALs. These systems tend to cluster in
velocity space within 5\,000\,\kms{} of the emission redshift
(often termed ``associated NALs''; Weymann \et 1979; Foltz \et 1986). 
The strongest indications that a
given NAL is associated with the quasar include: (a) time-variable
absorption strengths, (b) smooth well-resolved absorption profiles that
are broader than the thermal widths, (c) multiplet ratios implying
partial coverage of the continuum source, and (d) high particle densities
inferred from excited-state absorption lines (Barlow, Hamann,
Sargent 1997), when detected. Nonetheless, NALs can still be associated without showing
clear signs thereof (Hamann \et 1997). 

Early on, NALs were recognized to be key to the study of the tenuous,
non-emitting gas situated between the quasars and us, but the absorption 
in gas associated with the quasar is also important (a) as alternative and
independent probes of the central engine and the dynamics and structure
therein, and (b) for understanding how quasars evolve and affect their 
environment on small and large scales in the process.  Quasar intrinsic
NALs have been 
suggested to originate in winds driven off a torus (\eg Barthel, Tytler, \& 
Vestergaard 1997) or off the accretion disk (\eg Murray \et 1995; Murray
\& Chiang 1995; Elvis 2000; Ganguly \et 2001).  Some recent thoughts 
on possible origins of the NALs in context of the evolution of the central 
host galaxy as a result of either a recent merger or a burst of 
star formation are discussed by Hamann \et (2001), Baker \et (2002), and 
Sabra, Hamann, \& Shields (2002). Baker \et (2002) specifically argue that
dusty absorbing gas is closely related to the onset of radio activity.
At least for nearby low-ionization BAL quasars, there is evidence that the
BAL phenomenon may also be a temporary phase in the early evolution of the
quasar (or soon after its re-ignition; \eg Canalizo \& Stockton 2001, 2002).
A similar scenario has been proposed for the $z = 1.5$ radio-loud BAL quasar,
FIRST J1556$+$3517 (Najita, Dey, \& Brotherton 2000).  It is also possible 
that some (narrow) 
absorbers may originate in the host galaxy itself; this has been argued for 
at least  some Seyfert~1 galaxies (Crenshaw \& Kraemer 2001).
Many of the UV absorbers in Seyfert~1 galaxies also appear connected with
absorption at X-ray energies (\eg Mathur, Elvis, \& Wilkes 1995; Crenshaw \et 1999).
Such a connection between UV and X-ray absorbers is also seen in BAL quasars 
(Brandt, Laor \& Wills 2000; Mathur \et 2001; Green \et 2001) and a few 
non-BAL quasars (\eg Mathur \et 1994; Brandt, Fabian, \& Pounds 1996). 

It has long been debated whether or not the associated NALs and the BALs are related 
somehow (Weymann \et 1979) and how they may fit into the evolution of the 
quasars (Briggs \et 1984; Hamann \et 2001). A consensus is starting to appear 
that most associated NALs may be related to BALs, but the details of the 
connection differ (cf.\ Hamann, Korista,\& Morris 1993; Hamann \et 2001; 
Elvis 2000; Ganguly \et 2001; Laor \& Brandt 2002; see also Murray \& Chiang 
1995, 1997).  If BALs and associated NALs 
are related, it is relevant to keep in mind that BAL outflows most likely are 
near-equatorial (\eg Weymann \et 1991).  The most convincing evidence for 
this are likely polarization observations (\eg Ogle et al.\ 1999; Schmidt 
\& Hines 1999) combined with (a) estimates of the position angle of the radio 
source symmetry axis in a few nearby BALs quasars (Goodrich \& Miller 1995) 
and (b) covering fraction arguments (Hamann \et 1993). 

The first step toward an in-depth understanding of the issues of the origin
of the quasar-associated NAL gas, its connection to BAL outflows, and its 
importance as a probe
of the central engine is to know the quasar absorption frequency, and how this 
frequency and the NAL and BAL properties relate to the emission properties of 
the quasar, if any; intervening systems will be indifferent to the intrinsic 
quasar properties (see \S~\ref{associd}).
A high fraction (50\% $-$ 70\%) of low-redshift Seyfert~1 galaxies are absorbed, 
but BALs are not detected in these sources (\eg Crenshaw \et 1999; Hamann 2000).  
BALs are detected almost exclusively in the more luminous radio-quiet 
quasars (RQQs) at a frequency of 10\% $-$ 15\% (\eg Foltz \et 1990; 
Hamann \et 1993; Hewett \& Foltz 2003). Relatively few 
radio-loud quasars (RLQs) are so far known to display BAL systems (\eg 
Becker \et 1997, 2000, 2001; Brotherton \et 1998), 
although in most cases these objects may only appear mildly radio-loud 
owing to the absorbed optical emission yielding an artificially enhanced 
radio-loudness (\eg Goodrich 2001). But BAL troughs have been detected in a 
few bona-fide (\ie very radio powerful) RLQs (Gregg \et 2000; Najita \et 2000).

Unfortunately, the frequency of associated NALs among quasars is poorly known, 
especially among RQQs. While NALs have been studied almost as long as
quasars themselves (Perry, Burbidge, \& Burbidge 1978; Weymann, Carswell,
\& Smith 1981), a lot of the focus with respect to associated NALs has been 
on RLQs\footnote{
This may partly be due to the combination of (i) the early debate over the
reality of associated absorbers (Weymann \et 1979; Foltz \et 1986; cf. 
Young, Sargent, Boksenberg 1982; Sargent, Boksenberg, Steidel 1988), (ii)
the fact that larger samples of radio-selected quasars were available
for study early on, and (iii) the realization that strong NALs were
somewhat common in RLQs (see Foltz \et 1988 for discussion and review).
} 
and, even so, a proper census has not been made to date.  Well-defined, 
near-complete samples provide the best basis for a proper investigation 
of the associated-absorption frequency of quasars with different emission properties. 
Such samples and their spectral data are beginning to accumulate 
[\eg the Large Bright Quasar Survey (Hewett, Foltz, \& Chaffee 1995); 
the FIRST survey (Gregg \et 1996; White \et 2000); the Molonglo 
Quasar Survey (Kapahi \et 1998; see Baker \et (2002) for an absorption
study); and the upcoming releases of statistically well-defined and 
near-complete quasar samples detected in the Sloan Digital Sky Survey 
(\eg Schneider \et 2002) will undoubtedly be valuable].
Results on associated NALs in the UV are starting to appear, but mostly for low 
redshift quasars (Jannuzi \et 1998; Ganguly \et 2001; Laor \& Brandt 2002) 
and for more distant RLQs (Baker \et 2002).  

The main goal of this paper is to provide what appears to be the 
first assessment of the relative 
\civ{} \lam 1549 absorption frequency of RQQs compared to RLQs for moderate 
redshift ($z \approx 2$) quasars\footnote{While Richards \et (1999) and 
Richards (2001) also study moderate redshift quasars, they only study the 
{\it absorbed} quasars and therefore cannot assess how commonly NALs 
occur among various quasar subtypes.}.  
While the sample analyzed here is neither complete nor homogeneous, it does 
have the advantage of being of statistically significant size and being 
well selected to reduce luminosity and redshift effects, as is necessary 
for a proper comparison of the two radio-types (\S~\ref{smpls}).
Hence, the absorption frequencies, presented here, should provide 
reasonable guideline values until frequencies based on larger and 
more complete quasar samples appear.
It is also examined how the UV narrow absorption properties relate to 
UV and radio emission properties for clues to conditions conducive to 
these absorbers for different quasar types.
In particular, the current data suggest (see \S~\ref{results}) that the 
relatively strong associated NALs may be the low-velocity equivalents 
to the dramatic BALs.
This work is an extension of the preliminary results presented earlier 
(Vestergaard 2002; hereafter Paper~I), where results for the combined 
sample of RLQs and RQQs were mainly presented. Here, special attention 
is paid to the differences in absorption properties between the two quasar 
radio types and between the RLQ subtypes.  

The paper is organized as follows: \S~\ref{smpls} describe the selection
of the quasar samples, \S~\ref{data} details the data and spectral measurements,
while the results of the analysis are presented and discussed in \S~\ref{results}.
The results are discussed in context of recent relevant studies in 
\S~\ref{radioevolution} and \S~\ref{fwhm}. A summary and the conclusions are
presented in \S~\ref{conclusion}.

A cosmology with $H_0$ = 50 ${\rm km~ s^{-1} Mpc^{-1}}$, q$_0$ = 0.0, and 
$\Lambda$ = 0 is used throughout, unless otherwise noted.

\section{Quasar Samples \label{smpls}}

The quasar sample under study comprise 66 RLQs and 48 RQQs located at 
redshifts between 1.5 and 3.5;
it is studied already by Vestergaard (2000), Vestergaard, Wilkes, \& 
Barthel (2000), Vestergaard (2004a), and Paper~I. 
A forthcoming paper will present the details of the data acquisition and 
processing, and the rest-frame UV spectra. 
The details relevant for this work are briefly summarized below and 
in \S~\ref{data}.

There are several reasons why this quasar sample is particularly useful for 
this investigation. First, the objects were selected for a study of the
emission line profiles and, hence, were {\it not} selected on account 
of the presence of absorption lines in their spectra; the one exception is
the deliberate omission of known broad absorption line quasars since the
absorption can inhibit accurate measurements of the emission lines.
This selection allows for an 
assessment of the frequency of narrow absorbers in general and with respect to 
radio type. Second, the quasars in the two radio categories were selected to 
match one-on-one in luminosity and redshift to within reasonable measurement 
uncertainties of the absolute $V$-band magnitude, $\sigma(M_V) \approx 0.5$ 
mag and of redshift, $\sigma(z) \approx 0.01$.  
The RQQs and RLQs populate the same range in $M_V$ with an average 
$<M_V>$ = $-$27.9 mag, as illustrated in \S~\ref{absprops}. 
The pair-matching is used here only to ensure that the luminosity and 
redshift effects are minimized between the RLQs and RQQs, so not to 
complicate the analysis and the comparison of the two radio-subsets.  
Third, extensive databases of \civ{} broad emission line measurements 
(\eg Vestergaard 2000), and of good radio data of the RLQs (see below) are 
available.  Finally, all the rest-frame UV spectra are of relatively high-quality 
and are uniformly processed and measured (with reliable error estimates). 
Assessments can hence be made whether or not the observed absorption 
properties are related to UV emission properties of the quasars.

The RLQs were selected on the basis of unification models for extragalactic 
radio-sources to span a range in source inclination, $i$, with 
respect to our line of sight but such that the broad-line region is still 
visible ($i \gsim$44\degs; \eg Barthel 1989). This is roughly obtained by 
selecting individual RLQs with a
range in fractional radio-core dominance, estimated by $\log R_{\rm 5GHz} 
= \log [S_{\rm 5 GHz, core}/S_{\rm 5 GHz, total}]$ and by $\log R_{\rm V}
= \log [L_{\rm 5 GHz, core}/L_{V}]$ (\eg Wills \& Brotherton 1995), where
$S_{\rm 5 GHz}$ is the restframe 5\,GHz radio flux, $L_{\rm 5 GHz}$ is the
luminosity at 5\,GHz, and $L_{V}$ is the optical rest-frame $V$-band
continuum luminosity.  RLQs are sub-classified as lobe-dominated quasars, LDQs, 
($R_{\rm 5GHz} < 0.5$), core-dominated quasars, CDQs ($R_{\rm 5 GHz} \geq$0.5), 
compact steep-spectrum quasars, CSSs ($R_{\rm 5GHz} < 0.5$ and the largest 
linear size\footnote{The linear size constraint for a steep spectrum source 
to be called `CSS' is not uniform in the literature. CSSs are defined here 
to have a radio source extent less than 25 kpc in a cosmology with
$H_0$ = 50 ${\rm km~ s^{-1} Mpc^{-1}}$, q$_0$ = 0.0, and $\Lambda$ = 0 
to allow a direct comparison with the Baker \et (2002) study in 
\S~\ref{radioevolution}. } of the radio emission (LIN) is 25\,kpc or less in a 
cosmology with $H_0$ = 50 ${\rm km~ s^{-1} Mpc^{-1}}$, q$_0$ = 0.5, and 
$\Lambda$ = 0; see footnote~4), and giga-hertz peaked spectrum quasars, GPS 
(compact [$\lsim$\,1\,kpc] sources with convex radio spectrum peaking between
500\,MHz and 10\,GHz; O'Dea 1998). 
The LDQs are commonly interpreted to be viewed at relatively larger inclination 
angles than the CDQs; CDQs are believed to be viewed almost face-on ($i \lsim$ 
5\degs{} $-$ 10\degs).  To avoid the possible effects of Doppler boosted
(\ie beamed) optical-UV continuum emission on the emission line properties 
the most strongly core-dominated quasars were selected against by omitting
the known highly variable objects (at optical and radio energies), \ie blazars.
Final selection criteria were imposed to ensure that the RLQs are 
accessible from the Very Large Array (VLA) and are observable at the 
Palomar Hale 5\,m telescope (\ie $V \lsim$ 19 mag). The RQQs were selected 
from Hewitt \& Burbidge (1993) to match the RLQs in $M_V$ and $z$ as described 
above.  Unfortunately, no inclination estimator is available for the RQQs, 
so a similar sub-distribution in $i$ can not be selected.

\section{Data \label{data}}

Rest-frame UV spectra for most of the RLQs are presented by Barthel, Tytler, 
\& Thomson (1990).  The remaining quasars were observed mainly at the 
Multi-Mirror Telescope (MMT).  The spectra commonly cover rest-frame 
wavelengths $\sim$1000\AA{} to $\sim$2100\AA{} with a spectral resolution 
of $\sim$3.5\AA{} to $\sim$5\AA{}. Shorter exposures 
of many of the Barthel \et quasars were repeated at the MMT (a) to 
ease the combination of the red and blue subspectra, (b) to extend these
spectra to the atmospheric cutoff ($\sim$3200\AA), and (c) to provide a
better flux calibration, which for some spectra is less accurate.
The spectra were not corrected for Galactic extinction. 
This correction is typically within the uncertainty in the absolute
flux calibration, since these data are not spectrophotometric.  High quality 
Very Large Array radio maps at 1.4\,GHz, 5\,GHz, and 15\,GHz are available for 
most of the RLQs (Barthel \et 1988; Lonsdale, Barthel, \& Miley 1993; 
Barthel, Vestergaard, \& Lonsdale 2000) and fluxes are available for
many of the quasars at 178\,MHz or 159\,MHz (Spinrad \et 1995) and 408\,MHz
(Colla \et 1970, 1972, 1973; Large \et 1981; Wright \& Otrupcek 1990).

\subsection{Spectral Measurements\label{msmts}}

A power-law continuum ($F_{\lambda} \propto \lambda^{\alpha_{\lambda}}$) 
was fitted to virtually line-free regions in the 
spectra (see \eg Vestergaard \& Wilkes 2001).
For estimates of the continuum uncertainty and its effect on other
spectral measurements, the continuum level and slope were varied to
the four extremes permitted by the noise in the data (\ie the minimum, 
the maximum, the bluest slope, and the reddest slope). For each of 
these five continua the \civ{} emission line shape was reproduced with a 
smooth fit (using 2 -- 3 Gaussian profiles), which eliminates noise spikes 
and narrow absorption lines superposed on
the emission line profile. This smooth profile fit to the best fit continuum
is very suitable as the local ``continuum'' level for the absorption lines.
The emission line width, FWHM(\civ), (used in \S~\ref{fwhm}) was measured 
directly on this smooth profile, reducing measurement errors. The FWHM 
uncertainty was estimated from the FWHM measurements of the profile fit
made to the four extreme continuum settings.

A semi-automated absorption line search algorithm was employed, but all the
data and measurements were visually inspected and confirmed.
First, an extensive list of candidate \civ{} $\lambda \lambda$1548,1550 
absorption line positions was interactively generated from the spectra in 
the observed frame.  To minimize omissions, candidate absorbers with line 
peaks deviating more than $\gsim$1.5$\sigma_{\rm rms}$ ($\sigma_{\rm rms}$ 
is the local spectrum {\it rms} noise) from the local continuum were then 
selected; most candidates with smaller peaks are clearly noise.  This 
interactive search eliminates the predominant part of the noise spikes, 
which the human eye can more easily identify than an algorithm.  Thereafter,
the algorithm searches among these pre-selected candidate $\lambda$1548 and
$\lambda$1550 transitions for \civ{} doublet 
lines satisfying the following criteria:

\begin{enumerate}
\item
The peak of the $\lambda 1550.77$ transition, at least, is greater than 1.5$\sigma_{\rm rms}$.
\item 
The {\it observed} equivalent width $EW_{\rm obs}$ $\geq$ 0.5\AA{} of each system/blend
\item
The measured restframe separation of candidate doublet lines have \\
$\Delta \lambda_{\rm separation,rest} = \Delta \lambda_{\rm doublet} \pm \half$
resolution element, where the \civ{} doublet \\ separation is
$\Delta \lambda_{\rm doublet}$\,=\,1550.77\,$-$\,1548.20\AA{}\,=\,2.57\AA.
\item
Rest $EW$ doublet ratio ($\lambda 1550/\lambda 1548$) is in the range: 0.8 $-$ 2.2 
\item
Rest $EW$ of each doublet is at least 3 times the detection limit, $\sigma_{\rm det}$
\item
The doublet FWHMs generally match to within the resolution ($\sim$200$-$300 \kms)
\end{enumerate}

The FWHM criterion is not strictly enforced due to resolution/blending effects 
and measurement uncertainties; extensive tests show that this does not seriously 
affect the list of absorption candidates. Besides, the final visual inspection 
of the absorbers identifies clearly unrealistic or unreliable absorbers, which
may have survived the filtering. 
The margins in $\Delta \lambda_{\rm separation,rest}$ (item 3) and $EW$ doublet 
ratio (item 4) were adopted to allow for blending and resolution effects.
The imposed value ranges were determined from spectra with clear presence of 
\civ{} narrow absorption doublets.  The $EW$ detection limit ($\sigma_{\rm det}$) 
is determined as the integrated local continuum rms noise (normalized to the local 
continuum level) across a single resolution element. This yields an absolute 
minimum detection limit.  The $EW$ is measured by integrating across the 
identified absorption profile of the \civ{} doublet.  This is done because 
the spectral resolution and relative weakness of some of the absorption 
lines do not warrant employment of Gaussian fitting for line measurements.
Some of the absorption lines are clearly blended at the resolution of these 
data and they may appear as a single absorption line.  Therefore, such strong 
absorption blends, commonly with
$EW_{\rm rest} >$\,1.0\AA{}  and easily identified by eye, were not 
subjected to the above-listed selection criteria.  The search for NALs
were deliberately stopped at the restframe wavelength $\sim$1420\AA{} to
avoid the increasing probability of contamination by \siiv{} absorbers 
and other lower-redshift, intervening galaxy and/or cluster absorption.
Notably, criterion 5 above ensures that each \civ{} NAL is detected with
a confidence $\geq 3 \sigma$ (\ie the confidence in the detection is at
the $\geq$ 99.95\% level). The completeness level is determined in \S~\ref{compltness}.

\subsection{Identification of Associated Systems 
\label{associd}}

Unfortunately, for most of the objects (except Q1726$+$344; see below) it 
cannot be firmly established that a given NAL is associated with the
quasar based on the criteria listed in the introduction, since only 
single-epoch, moderate resolution data are available.  
Ideally, intervening \civ{} absorbers can be identified by 
searching for absorption lines by other species, such as \mgii, at
the same redshift of the individual \civ{} absorbers.  This is only possible
in a very limited number of objects for which \mgii{} appears in the spectral
window.  
Fortunately, most of the intervening absorption lines are anticipated
to be relatively weak and therefore to be eliminated by the primary 0.5\,\AA{} 
$EW_{\rm obs}$ cutoff. Savage \et (2000) find the strongest \civ{} absorption 
lines due to Milky Way gas detected in quasars to have $EW_{\rm obs} \lsim$0.9\AA.  
The likelihood that any of the strong absorption systems
($EW_{\rm rest} \geq$ 1\AA) are intervening is considered to be rather small.
Hence, it is fair to assume that the stronger an absorption line is, the more 
probable is its association; the discussion in \S~\ref{results} confirms this.
Note that not all the weak absorbers (\eg with $EW_{\rm rest} <$ 1\AA) are per se
intervening. A characteristic property of unrelated, intervening systems to
bear in mind is that they should scatter evenly in redshift (up to the
emission redshift) and they do not typically exhibit trends with the quasar 
properties\footnote{
A possible exception is that of dusty damped Lyman-$\alpha$ absorption
(DLA) systems with detectable metal lines, which will redden the quasar 
spectrum.  It may appear that intervening dusty DLAs can lead one to
erroneously conclude that the frequency and strength ($EW$) of absorbers increase
with dust reddening (as is, in fact, seen for this sample; discussed in 
\S~\ref{results}). However, this is not very likely. 
The reasons are that (1) the incidence of DLAs is very low (by a factor $\sim$0.04;
Table 12.1 by Peterson 1997) compared to \civ{} narrow absorbers (intrinsic and 
intervening), and (2) DLAs have rather low metal abundances, resulting in 
typically weak \civ{} absorbers ({\it observed} equivalent width $\lsim$1\AA; 
\eg Prochaska \et 2003). Therefore, very few, if any, intervening \civ{} NALs
associated with DLA systems are expected in the current sample of absorbers 
($<$4 absorbers) which will not adversely affect the results of this study.
}.

\setcounter{figure}{0}

\begin{figure*}[t]
\vspace{-1.70cm}
\begin{center}
\hspace{-0.3cm}
\hbox{
\vbox{
\hbox{ 
\epsfxsize=9.0cm
\epsfbox{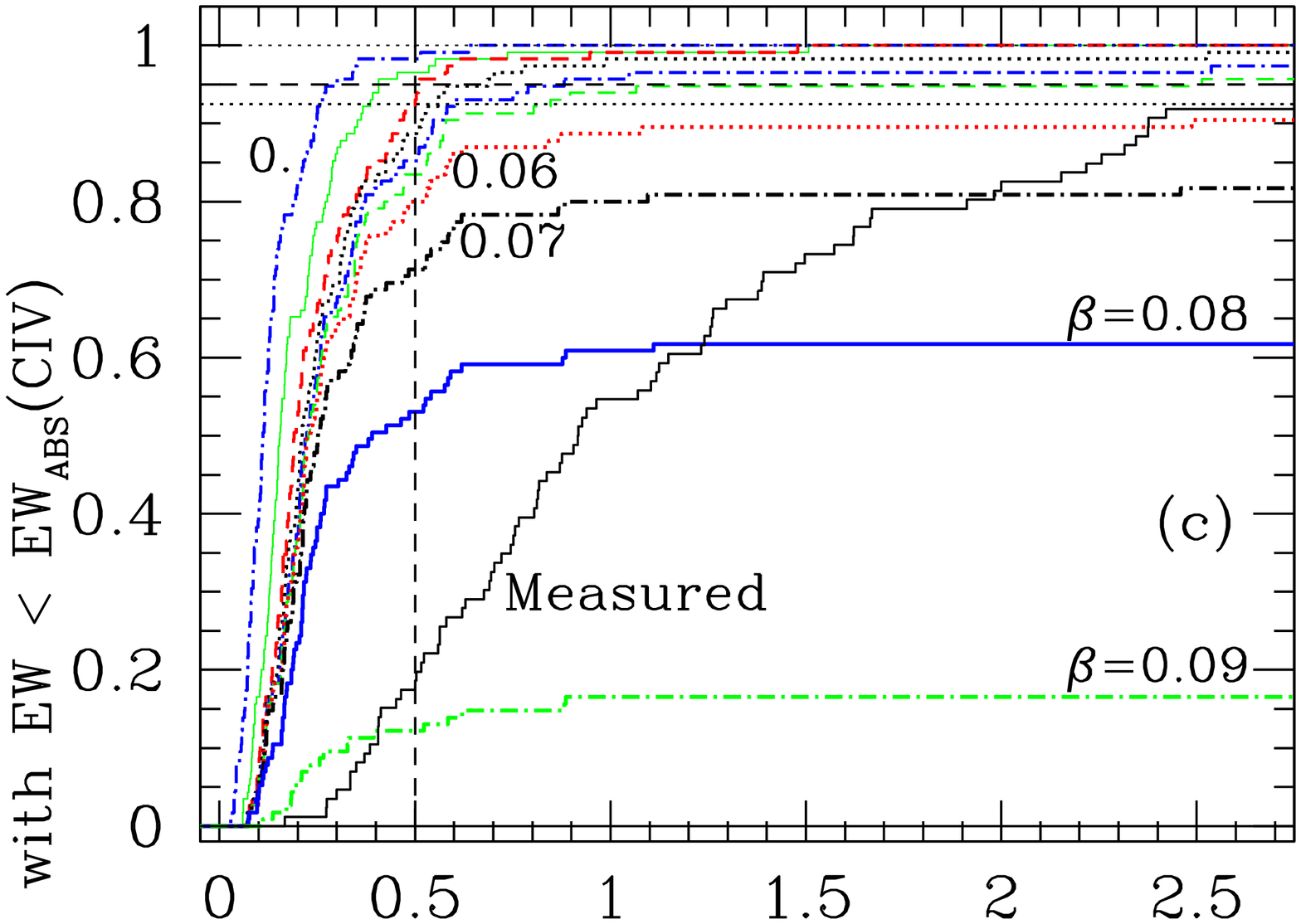}
}\vspace{-2.5cm}\hbox{
\epsfxsize=9.0cm
\epsfbox{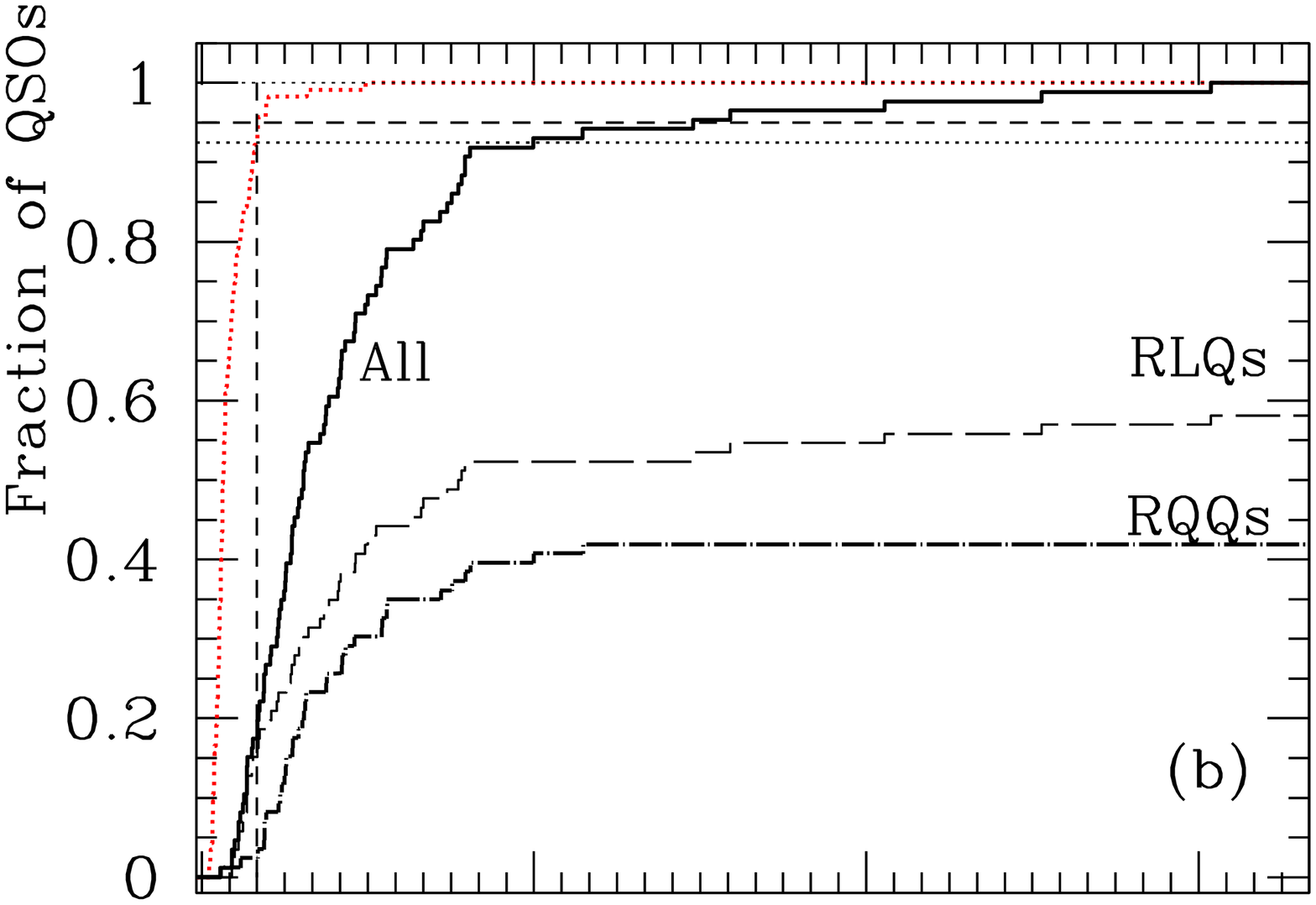}
}\vspace{-3.27cm}\hbox{
\epsfxsize=9.0cm
\epsfbox{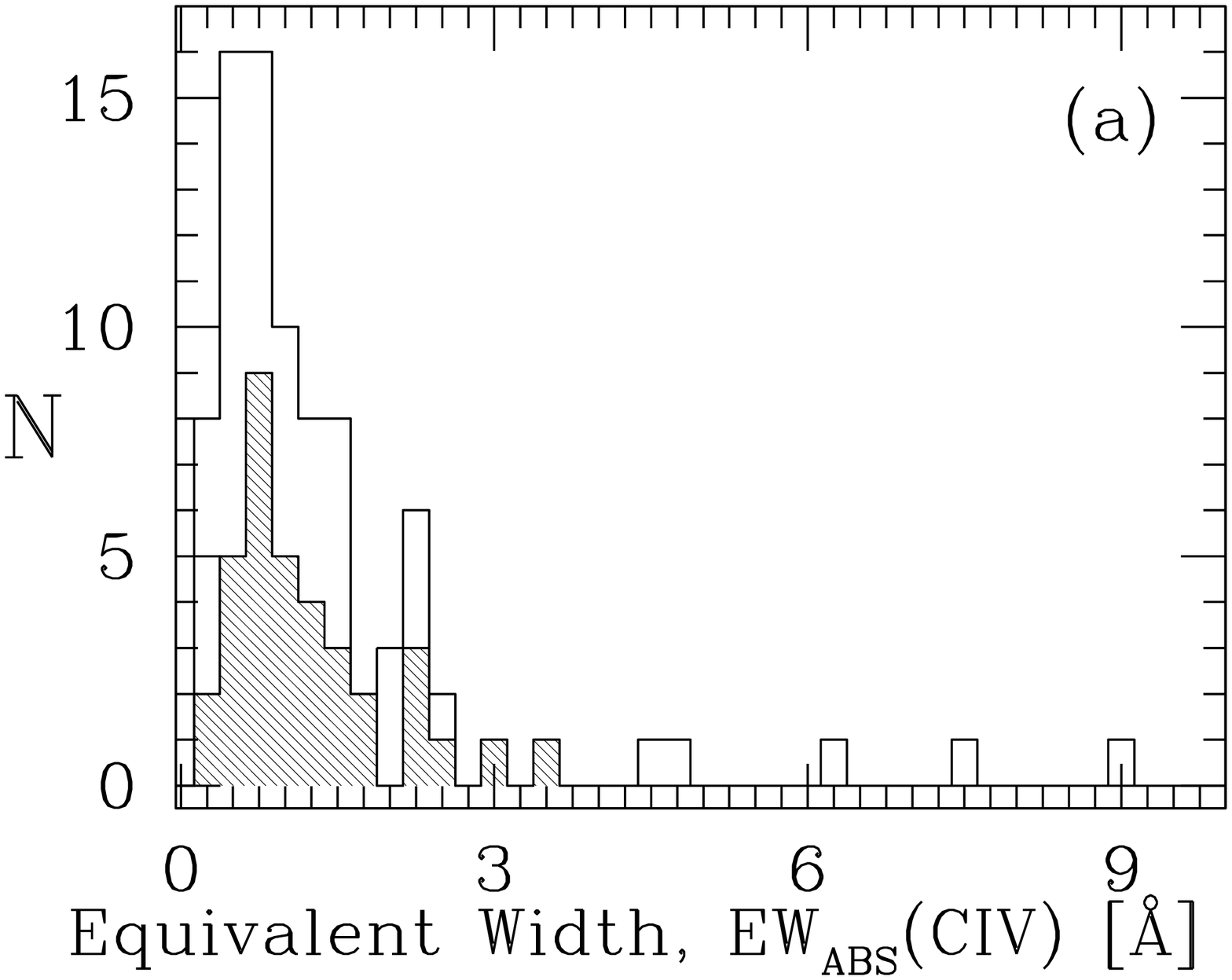}
}}\hspace{0.2cm}\vbox{ 
\hbox{ 
\vspace{5.5cm}\hspace{0.2cm}
\begin{minipage}{8.25cm}
\caption[]{ Distributions of restframe $EW_{\rm abs}$(\civ). 
($a$):  Restframe $EW_{\rm abs}$ distribution of the \civ{} doublet for the 
sample of absorbed quasars (solid open histogram); the RQQ subset is shown shaded.
($b$): Cumulative distribution of the measured $EW_{\rm abs}$(\civ)
(heavy solid curve) is compared with the cumulative distributions of the NALs
detected in the
RLQs (long dashed) and RQQ (dot-dashed) subsamples. The 3$\sigma$ detection limit
for $\beta = 0.02$ (heavy dotted curve; see below) is shown for comparison. 
Vertical and horizontal lines are as in panel (c).
($c$):  Cumulative distribution of the 3$\sigma$ restframe $EW$ detection limit for
different $\beta$ values. The leftmost curve (dot-dashed) is the distribution for
$\beta = 0.0$. Going from left to right the $\beta$-value increases by 0.01. The 
distributions for $\beta = -0.01, -0.02$ are essentially the same as those for
$\beta = 0.01, 0.02$, respectively, and are therefore not shown for clarity.
However, the completeness limits for $\beta < -0.02$ stay rather high (see \eg
Fig.~\ref{vel.fig}) contrary to those shown for $\beta > 0.02$.
The cumulative distribution of the measured $EW_{\rm abs}$(\civ) of all the quasars
is shown for comparison (labeled `Measured'). 
The 92.5\% and 95\% completeness levels are marked by the dotted and dashed 
horizontal curves, respectively. The 3$\sigma$ detection limit measured (in
a resolution element) to a 
completeness level of $\sim$95\% within $|\beta| < 0.02$ is $EW_{\rm rest} = 
0.5$\AA{} (dashed vertical line).
\label{ewdistrib.fig}}
\end{minipage}
}}}
\end{center}
\vspace{-0.5cm}
\end{figure*}

Barthel \et (1990) identified a number of absorption systems of different
species in the spectra of their RLQs. While all the narrow \civ{} absorption
lines are measured independently in this study, a cross-check was made with
the Barthel \et results to eliminate the intervening absorbers identified
by them; only a few (six) such absorbers were to be eliminated (corresponding 
to 6.5\%$\pm$1.5\% of the originally identified systems). The 
\civ{} systems with identified UV transitions of, for example,
\mgi\,\lam 2853, \mgii\,\lam \lam 2796, 2804, \alii\,\lam 1671, 
\siiv\,\lam \lam 1394, 1403, and/or some of the many UV \feii{} and \siii{}
transitions (\eg Barthel \et 1988; Savage \et 2000)
at the same absorption redshift were considered to be intervening. 
The two high-velocity NALs of Q2222$+$051 and Q2251$+$244 are part of systems 
with additional \alii, \aliii \lam 1857, \siiv, and/or \cii{} transitions, but are here 
considered to be intrinsic given their large strengths, $EW_{\rm rest} >$ 1\AA.
The \civ{} absorbers with \nv{} $\lambda$ 1240 at the same absorber redshift 
must be associated with the quasar given the high level of ionization
(\eg Weymann \et 1979; Hamann 2000; Savage \et 2000). 
(Notably, while high-density absorbers may display mostly low-ionization 
transitions, this ionization cut will not significantly select against 
such absorbers, since they also tend to display strong lines with $EW_{\rm rest} \gsim$ 1\AA).
Hence, the incidence of intervening absorbers should be reduced for 
the RLQs in this analysis.  
If the RLQ fraction of intervening absorbers is typical also for RQQs, then
4 of the NALs detected in the RQQ spectra (or 5\% of the entire NAL sample)
are statistically expected to be intervening (but not identified as such).

While a few intervening systems may remain, it is anticipated that their 
inclusion will not significantly affect the results of the analysis, 
because (a) the samples of quasars and absorbers are large, and 
(b) this study is of a statistical nature. 
In addition, even if the ensemble of the weak absorbers (or bulk thereof) 
are unrelated to the quasars their properties (\eg frequency,
strength, velocity) are expected to be indifferent to varying quasar
emission properties anyway, as noted above (see also footnote above).
In the following, it will be assumed that the narrow absorbers are intrinsic
to or associated with the quasars, unless otherwise noted.

For one RLQ, the CSS quasar Q1726$+$344, Ma (2002) detected a newly 
developed mini-BAL; only intervening absorption was detected here. Variable 
absorption is a clear indication that the absorber is associated 
with the quasar (\S~\ref{intro}).
The rest $EW$ = 2\AA{} and velocity shift = $-$6000\,\kms{}, measured by 
Ma (2002), are thus adopted in this work for comparison.

\subsection{Absorber Measurements and Completeness Limits \label{compltness}}

In the following, all $EW$s are restframe measurements
of the \civ{} absorption doublet. Table~\ref{dataobsabs.tab} lists the UV 
properties of the absorbed quasars and their absorbers, and 
Table~\ref{dataObsNoAbs.tab} lists the basic properties of the unabsorbed objects.
Figure~\ref{ewdistrib.fig} shows the $EW$ distributions of the individual \civ{} 
absorbers and the spectral detection limits, as explained next.

It is evident from Figure~\ref{ewdistrib.fig}a that while
the absorbers span a large range in strength, most have $EWs \lsim 3$\,\AA. 
The NALs in RLQs and RQQs generally distribute similarly, with the noticeable 
exception that only RLQ NALs are detected at $EW > 4$\AA{}.

In Figure~\ref{ewdistrib.fig}b the cumulative distribution of detected NALs 
is compared to the individual distributions of RLQ and RQQ absorbers. 
The dotted curve represents the cumulative distribution of the 3\,$\sigma$
detection limit of NALs within 6000\,\kms{} of the emission line redshift,
and shows a 95\% completeness level of NALs with $EW \geq$ 0.5\AA{} (see below).
Evidently, only a tiny fraction of the RQQ absorbers are below the 95\% 
completeness limit.

Figure~\ref{ewdistrib.fig}c shows the cumulative distributions of 
the 3\,$\sigma$ $EW$ detection limits measured for different absorber
velocities (parameterized by $\beta$) in the spectra of all absorbed and
unabsorbed quasars. The absorber velocity is conventionally defined as
$v = \beta c$, where $\beta =(r^2 - 1)/(r^2 + 1)$ and 
$r = (1 + z_{\rm em})/(1 + z_{\rm abs})$ (Weymann \et 1979; Peterson 1997) 
and $c$ is the speed of light. However, in this work $v = - \beta c$
is adopted such that a negative velocity (intuitively) corresponds to
blueshifted absorption velocities. Only distributions for positive
$\beta$ values are shown in Figure~\ref{ewdistrib.fig}c for clarity. The
distributions for $\beta = -0.01, -0.02$ are essentially the same as 
those for $\beta = 0.01, 0.02$, respectively. 
The cumulative distributions show that the sensitivity of the data
toward detecting NALs (at the 3\,$\sigma$ significance) within 
$|\beta| \leq$ 0.02 drops below 95\% at $EW \approx 0.5$\AA{} 
(vertical dashed line). In other words, associated NALs with
$EW \geq$ 0.5\AA{} are $\gsim$\,95\% complete.  
For increasing $\beta$-values, the cumulative distributions shift 
to higher $EWs$ and lower fractional completeness.
A similarly high level of completeness cannot be achieved at all $\beta$ 
values, partly owing to differences in signal-to-noise levels.  
Furthermore, at the highest $\beta$ values ($\beta \geq 0.06$) the 
distributions do not reach 100\% because the spectra could not all be 
measured at these $\beta$ values. More specifically, at $\beta = 0.08$ (0.09), 
only $\sim$62\% ($\sim$17\%) of the spectra could be measured.
Nevertheless, the completeness level of $EW \geq$ 0.5\AA{} absorbers is 
still high ($\geq$70\%) for $\beta \leq 0.07$. The $\beta < 0$ cumulative 
distributions maintain a rather high completeness ($>$\,85\% for 
$-0.09 \leq \beta \leq 0$), contrary to the distributions for $\beta > 0.02$.
The completeness level of $EW \geq$ 0.5\AA{} NALs as a function of $\beta$ is 
discussed in relation
to the measured absorbers in \S~\ref{veldistrib}. Part of the cumulative
distribution of detected absorbers is also shown in Figure~\ref{ewdistrib.fig}c
for comparison. This shows that more than 80\% of the absorbers 
have $EW$s above the 0.5\AA{} completeness limit.

\section{Results and Discussion \label{results}}

As described in \S~\ref{intro}, a basic census of the associated-absorption 
frequency among distant quasars is poorly known, particularly among RQQs. Also,
little is known about which quasar properties are favored or evaded for 
objects with strong (associated) NALs and how these properties compare for the
two radio types. In the following, these issues are addressed with
the current data. In addition, the dependence on radio source inclination
(\S~\ref{inclin}), and the velocity distribution of the NALs 
(\S~\ref{veldistrib}) are discussed.

\subsection{Frequency of Occurrence \label{freq}}

Table~\ref{freq.tab} lists the fraction of absorbed quasars grouped
in various ways, and Table~\ref{freq3.tab} contains the absorber
frequency among the absorbed quasars only.
As pointed out in \S~\ref{compltness} the most complete subset of
NALs are at $\beta \leq 0.07$ and have $EW \geq$ 0.5\AA. To allow
a more direct and quantifiable comparison with future
quasar absorption studies, most of the statistics in Tables~\ref{freq.tab}
and~\ref{freq3.tab} are therefore limited to this most complete
subset. If all absorbers are counted the frequencies would in places
increase by a few per cent. This is mostly within the counting errors.

In Tables~\ref{freq.tab} and~\ref{freq3.tab}, statistics are listed 
for various subdivisions of the full quasar sample (col.~1): 
``All QSOs'' (RLQs$+$RQQs), RQQs, RLQs, and the subgroups of RLQs: 
CDQs, LDQs, CSSs, and GPSs, defined in \S~\ref{smpls}; the size of 
each of these samples is in col.~2. Columns 3 to 5 list numbers and 
frequencies for various NAL velocity bins, as marked.
Table~\ref{freq.tab} is divided into three sections (top, middle,
bottom) according to the absorption strength. 
In particular, the statistics in the top section are based on all NALs 
with any detected $EW$ (col.~3, left) and on individual NALs with 
$EW \geq$ 0.5\AA{} (col.~3, right; cols.~4 and~5). 
Table~\ref{freq3.tab} is structured similarly to Table~\ref{freq.tab}, 
except that the statistics of weak absorbers are omitted.

When counting all detected absorbers (all $EW$s) more than half of the 
quasars have NALs (Table~\ref{freq.tab}; note that this fraction is 
$\gsim$60\% if all the intervening systems are included).
This is comparable to the frequency ($\sim$50 $-$ 70\%) observed among 
Seyfert~1 galaxies (Crenshaw \et 1999), and shows how common this 
phenomenon is.  This frequency is higher than that seen for the Bright 
Quasar Survey (BQS; $z < 0.5$) quasars studied by Laor \& Brandt (2002) 
(40\% NALs and $\sim$10\% BALs). 
About 25\% of the quasars have associated NALs ($EW \geq$ 0.5\AA; 
Table~\ref{freq.tab}; note, this also holds when restricted to $EW \geq$1\AA{},
the commonly adopted division between intrinsic and intervening absorption 
(\S~\ref{data})).  Ganguly \et (2001) find a similar frequency of 
associated NALs (15/59 $\approx$ 25\% $\pm$7\%) among their sample of 
$z \lsim 1$ quasars, while Baker \et (2002) find a very high fraction 
of associated NALs (with 0.3\AA{} $\leq EW \leq$ 5.2\AA) in their RLQ 
sample (50\% $\pm$17\% at $0.7 \leq z \leq 1.0$; 
90\% $\pm$21\% at $1.5 \leq z \leq 3.0$; see \S~\ref{radioevolution} 
for discussion of the Baker \et results).
However, in reality, a direct comparison of the incidence of absorbers
with the Ganguly \et study is non-trivial as there is only a small common range in
$EW$ of the detected absorbers between the two samples: the current sample
is incomplete at $EW <$\,0.5\,\AA{} and contain a large number of relatively
strong absorbers ($\gsim$1\,\AA), while the Ganguly \et absorbers are mostly
weaker than 1\,\AA{}. In the common $EW$ range from 0.5\,\AA{} to 1\,\AA{}
(summed per quasar) of associated (\ie low-velocity) \civ{} NALs, Ganguly
\et find 4 absorbed quasars out of 59, or $\sim$7\%\,$\pm$3\%. In comparison, 
the current sample has 4/114 quasars with such associated weak absorbers 
(\ie 3.5\%\,$\pm$\,1.7\%).
While this may appear to suggest a marginal difference in incidence, both 
samples are subject to small number statistics and strong conclusions based
thereon are cautioned.

The RQQs appear slightly less frequently absorbed than RLQs, but their
frequencies are consistent within the counting errors; this also holds when
considering the low and high velocity absorbers separately (Table~\ref{freq.tab}). 
Even if broad absorption lines (BALs) are also counted, these frequencies 
may not change much because the BAL fraction\footnote{ 
Among optically selected samples $\sim$10\% $-$ 15\% of the 
quasars are BALs (\eg Foltz \et 1990; Goodrich 2001; Hewett \& Foltz 2003). 
The FIRST survey has identified (Becker \et 2001) a similar fraction 
($\sim$10\%) of BALs among bona-fide RLQs (\ie the radio-loudness, 
$R^{\ast} = L_{\rm radio}$/$L_{\rm optical} >$ 10), although most of the 
detected BAL quasars are radio-intermediate sources ($ 1 \leq R^{\ast} \leq 10$).}
is relatively low for both RQQs and RLQs
and the counting errors are at the level of the BAL fraction 
or larger. The RLQs and RQQs only exhibit apparent differences in the frequency 
of weak, low-velocity (\ie associated) NALs. However, this may not be real
since a larger number of NALs with $EW$ below the 0.5\,\AA{} completeness limit 
occur among RLQs than among RQQs (Fig.~\ref{ewdistrib.fig}b).


\setcounter{figure}{1}
\begin{figure*}[t]
\hspace{1.25cm}
\vbox{
\vbox{
\epsfxsize=15.0cm
\epsfbox{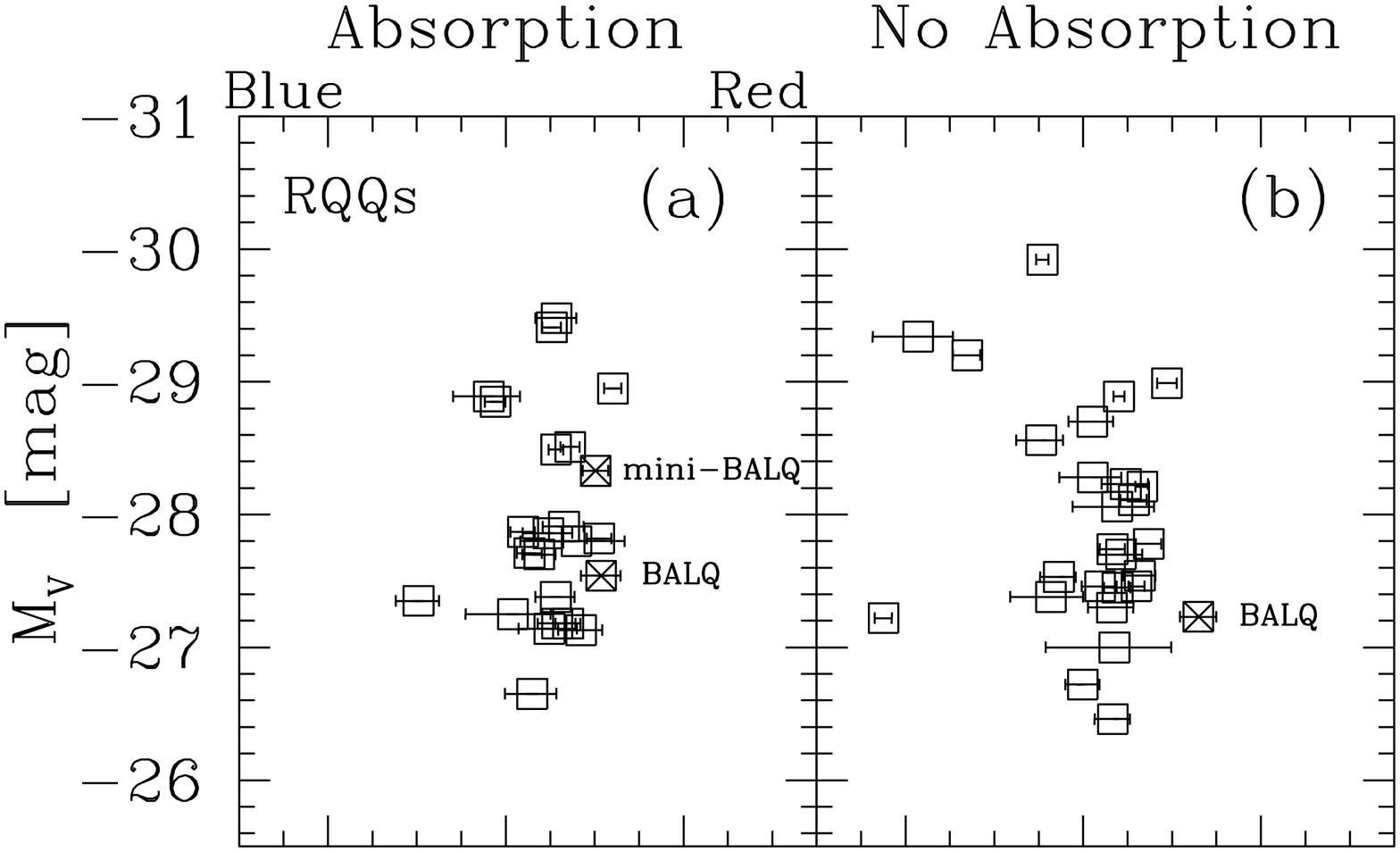}
}\vspace{-0.53cm}\vbox{
\epsfxsize=15.0cm
\epsfbox{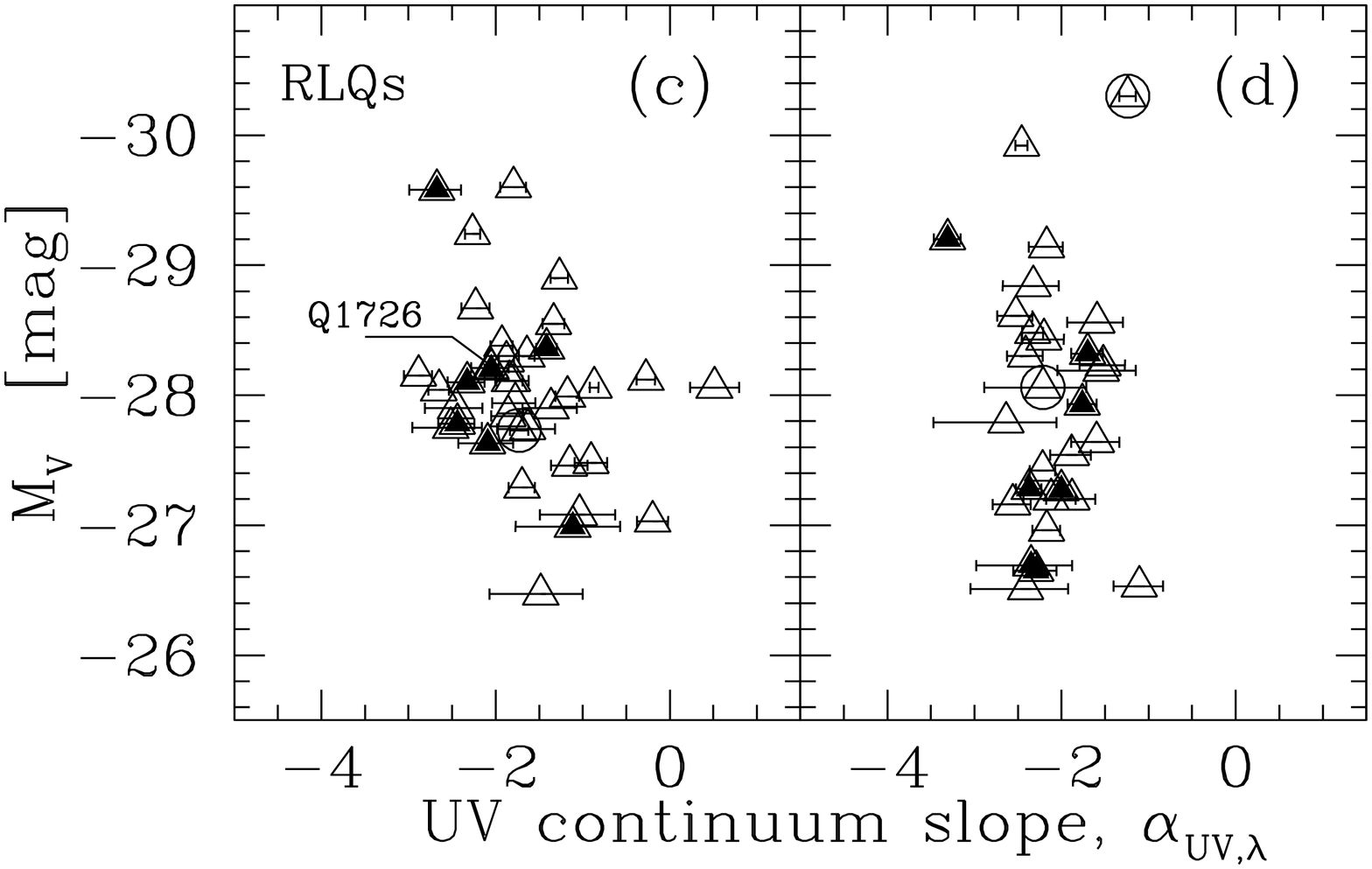}
}}
\vspace{-0.9cm}
\begin{center}
\caption[]{ Distributions of M$_{\rm V}$ and $\alpha_{\rm UV,\lambda}$
($F_{\lambda} \propto \lambda^{\alpha_{\lambda}}$)
for the quasars [($a$) and ($c$)] with \civ\ NALs (triangles: RLQs; squares: RQQs) and 
[($b$) and ($d$)] without \civ\ NALs. The typical uncertainty in M$_{\rm V}$ is $\sim$0.5 mag.
The RQQs with broad absorption lines (BALs) and mini-BALs 
(one BAL RQQ has no NALs and is hence plotted in panel b) and the RLQ Q1726$+$344 with a mini-BAL are  marked. 
The compact steep spectrum (CSS; solid triangles) and giga-hertz peaked radio 
sources (GPS; encircled triangles) are shown with separate symbols.
\label{mvalfa.fig}} 
\end{center}
\end{figure*}

Among the subtypes of RLQs, the CDQs and CSSs are typically equally
absorbed within the errors (Table~\ref{freq.tab}; top section). The 
LDQs stand out by having significantly higher occurrence of low-velocity
NALs (80\% $\pm$23\%) relative to CDQs and CSSs ($\gsim 20$\%). For the
strongly absorbed quasars there is a difference in absorption frequency
between the LDQs and CDQs at the $\sim 1\,\sigma$ level 
(Table~\ref{freq.tab}; middle section). This is seen both among the
low velocity absorbers ($v \leq$ 5000\,\kms; col.~4) and in the sample
of $v \leq 21\,000$\,\kms{} absorbers (col.~3). But, the high velocity
LDQ and CDQ absorbers are different at the $\sim 2\,\sigma$ level (col.~5).
Differences are also seen among the weakly absorbed LDQs and CDQs at the
$\geq 1.5\,\sigma$ level (col.~3; bottom section). Furthermore, Table~\ref{freq.tab} 
shows that CDQs tend to have more weakly absorbed objects than the
LDQs (col.~3). Most of this difference occur at high velocities (col.~5). 
The statistics for the weakly absorbed quasars should, however, be treated
with caution since the subset of $EW <$0.5\AA{} absorbers is incomplete 
(\S~\ref{compltness}).

It is evident from Table~\ref{freq3.tab} that most of the absorbers in 
any quasar subgroup are at low-velocity (\ie likely associated; col.~4), especially 
the strong absorbers. Again, the frequency of low and high velocity absorbers,
respectively, are similar between the RLQ and RQQ subgroups. This holds for 
the complete NAL sample ($EW \geq 0.5$\AA) and for the strong ones alike.  
Only the frequency of strong absorbers in LDQs and CDQs are different:
all strongly absorbed ($EW \geq$1\AA) CDQs have associated NALs only.

To summarize, RLQs and RQQs exhibit similar absorption frequencies 
both at high and low absorber velocity, respectively. Among the RLQs, 
the LDQs are by far the most frequently absorbed in general. Also, 
LDQs tend to be 
the most {\it strongly} absorbed. These strong absorption systems mostly 
have velocity shifts less than 5000\,\kms{} from the emission redshift. 
This will be discussed in section~\ref{inclin}. 
The CSS quasars are discussed in \S~\ref{radioevolution}.

\setcounter{figure}{2}

\begin{figure*}[t]
\hspace{1.25cm}
\vbox{
\vbox{
\vspace{-0.60cm}
\epsfxsize=15.0cm
\epsfbox{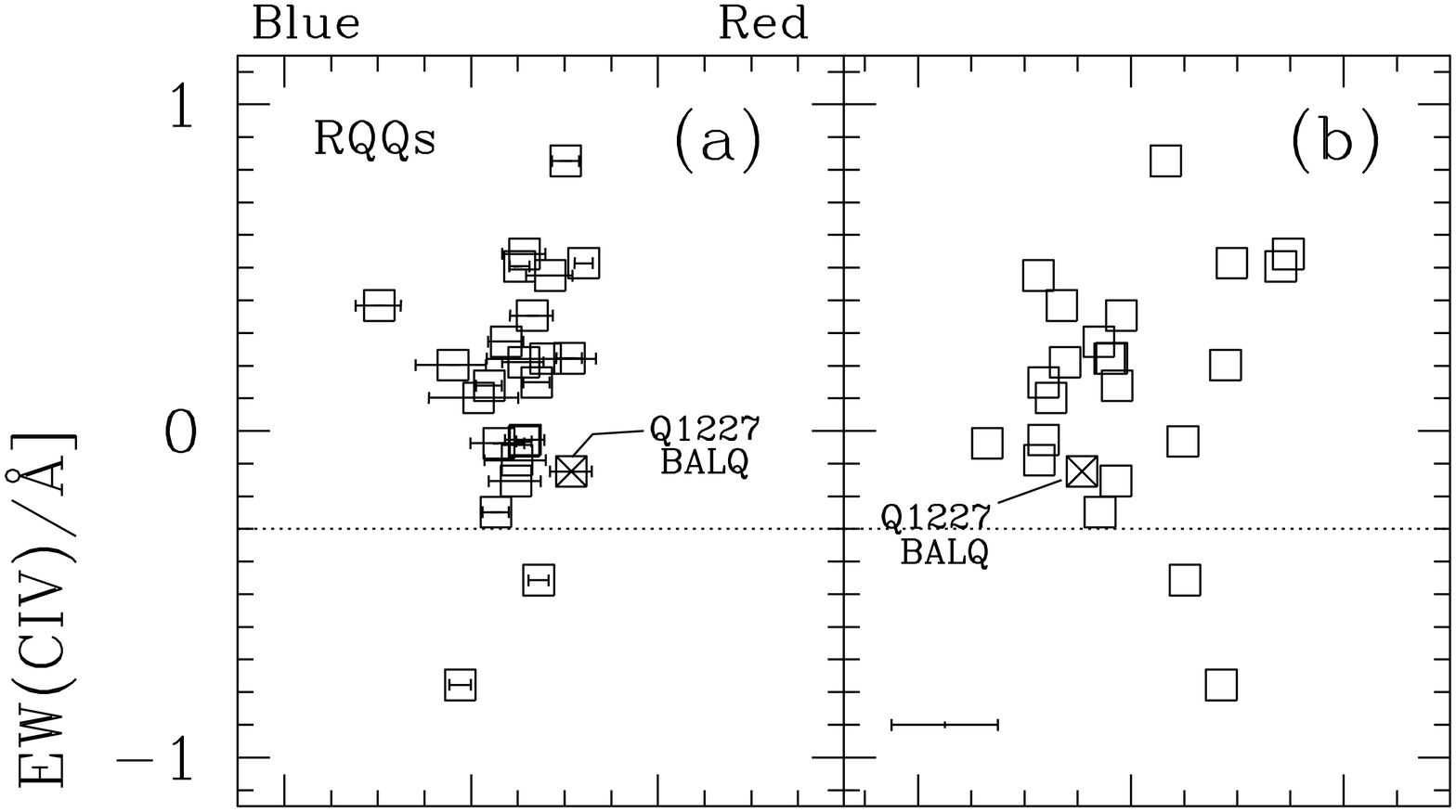}
}\vspace{-0.53cm}\vbox{
\epsfxsize=15.0cm
\epsfbox{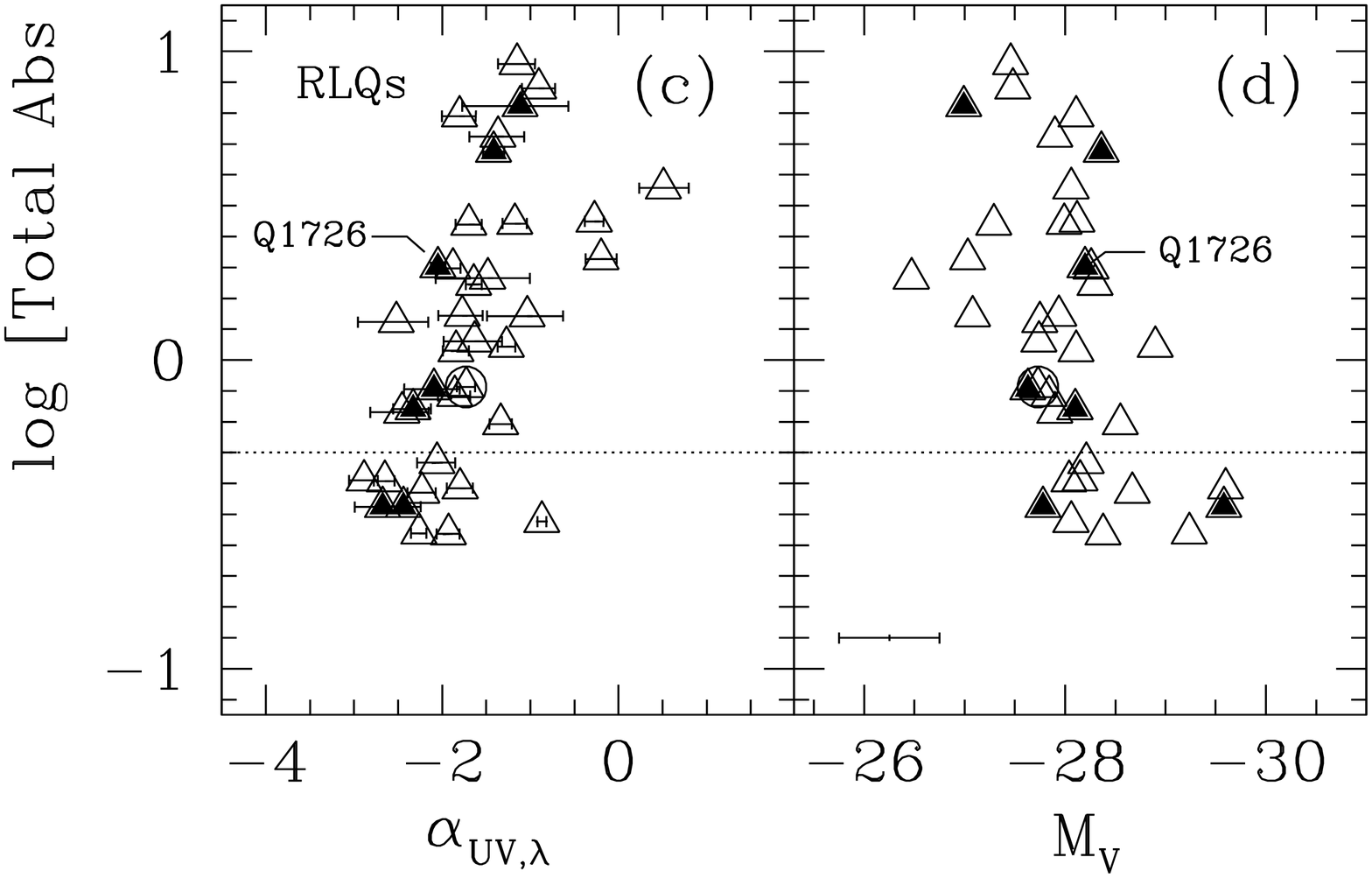}
}}
\vspace{-0.8cm}
\begin{center}
\caption[]{Distributions of the total restframe strength of the narrow absorbers ($EW$) 
in each quasar relative to ({\it left}) the UV continuum slope, $\alpha_{\rm UV,\lambda}$, 
and ({\it right}) to the quasar luminosity, M$_{\rm V}$.
The strengths of the narrow absorber detected in the broad absorption line quasar, Q1227$+$120, 
and of the RLQ Q1726 mini-BAL are highlighted. Symbols are otherwise as in Figure~\ref{mvalfa.fig}.
The dotted lines show the 95\% completeness level (0.5\AA).
\label{ewalfmv.fig}} 
\end{center}
\end{figure*}

A final comment is in place.
For a large, heterogeneous sample of absorbed quasars Richards \et (1999)
find marginal fractional excesses of \civ{} \lam 1549 NALs in (a) RQQs
compared to RLQs, and in (b) FSS ($\simeq$ CDQ) quasars relative to SSSs 
($\simeq$ LDQs), reaching velocities of $-$30\,000\,\kms{} and $-$25\,000\,\kms, 
respectively.  Richards (2001) confirms these findings to be marginal by 
including a more homogeneous data set of RLQs. Although the relative 
frequencies of high-velocity ($-$5\,000\kms{} to $-$21\,000\,\kms) absorbers 
seem to support the above-mentioned excesses (column (4) in 
Table~\ref{freq3.tab}, top), strong conclusions based thereon are,
unfortunately, premature for two reasons:
(1) this quasar sample is incomplete and the high-velocity absorption results
are based on small number statistics, especially for the RLQ subgroups, 
and (2) Richards \et (1999) use
an advanced Monte-Carlo type of normalizing algorithm in determining
the relative frequency of absorbers at a given velocity for each sample,
while a simple ratio is used here. The nature of the sample studied here
(item 1) does not warrant such an elaborate analysis.
The results of Richards \et (1999) and Richards (2001) will therefore
not be discussed further.
While the findings by these authors are indeed tantalizing, to 
properly address these issues, complete, homogeneous, and well-selected 
samples are required.

\subsection{Properties of Quasars with NALs \label{absprops}}

The distributions of UV continuum slopes, $\alpha_{\rm UV, \lambda }$ 
($F_{\lambda} \propto \lambda^{\alpha_{\lambda}}$) 
and the absolute magnitude, M$_{\rm V}$, are shown in Figure~\ref{mvalfa.fig}
for the quasars with and without detectable \civ{} $\lambda$ 1549 
narrow-line absorption, respectively.  ``Unabsorbed'' quasars (note, one BAL
quasar is included here) have on average a slope, 
\mbox{$<\alpha_{\rm UV}>$\,=\,$-$1.99} (median\,=  $-$1.89; standard deviation,
$\sigma_{\rm std}$ = 0.64), while the absorbed quasars in Figure~\ref{mvalfa.fig} 
have \mbox{$<\alpha_{\rm UV}>$\,=\,$-$1.59} (median\,=\,$-$1.62; 
$\sigma_{\rm std}$ = 0.64).  When confined to the absorbers above the 95\% 
completeness limit of 0.5\AA{} the absorbed quasars are slightly redder on 
average ($<\alpha_{\rm UV}>$\,=\,$-$1.47; median\,=\,$-$1.46; 
$\sigma_{\rm std}$ = 0.60).  Although the average slopes only differ at 
the 1\,$\sigma$ level, Kolmogorov-Smirnov (K-S) tests confirm that the 
continuum slope distributions of absorbed and unabsorbed quasars are 
statistically different at the 99.95\% confidence level or higher.
Note, how much more the absorbed quasars scatter in continuum slope,
in clear contrast to the unabsorbed sources; barring a few outlying 
unabsorbed RQQs.
Notably, most of the quasars are more luminous than $M_V = -27$ mag, inconsistent
with the prediction by M\o ller \& Jakobsen (1987) that no associated
absorbers should be seen for quasars more luminous than this value. About
25\% of these luminous quasars have strong ($\gsim$\,1\AA), associated 
NAL systems (Table~\ref{freq.tab}; middle section).

\begin{figure*}[t]
\hspace{1.25cm}
\vbox{
\epsfxsize=14.0cm
\epsfbox{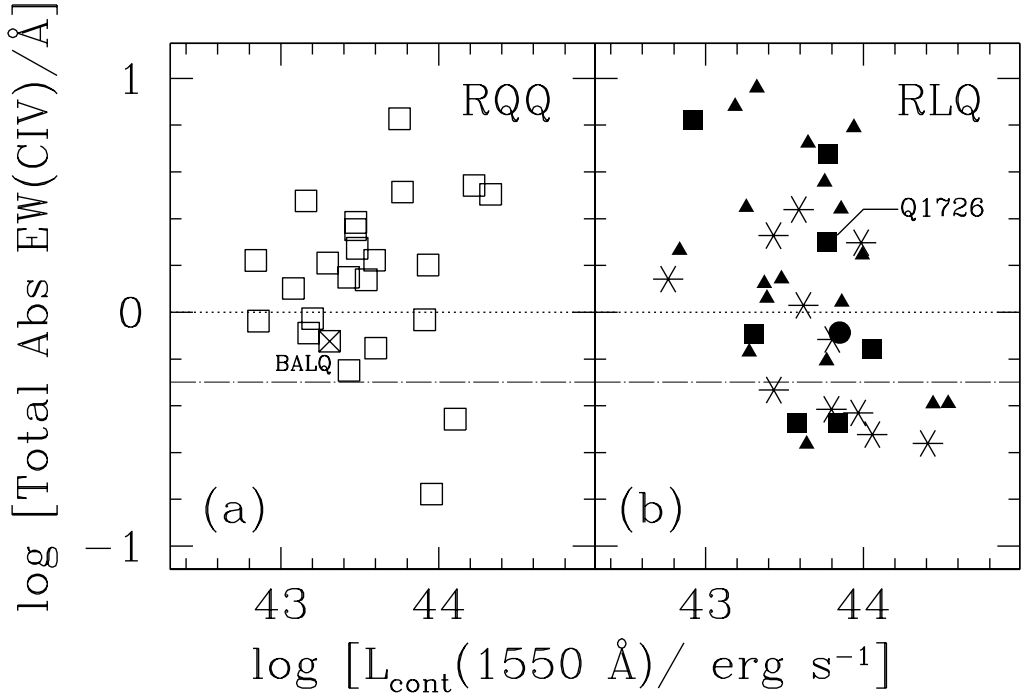}
}
\vspace{-0.5cm}
\begin{center}
\caption[]{Distributions of the total narrow absorber restframe strength, $EW$, for each quasar
with the UV continuum
luminosity for ({\it a}) the RQQs and ({\it b}) the RLQs. Symbols: open squares: RQQs;
solid squares: CSS; asterisks: core-dominated quasars (CDQ); small solid triangles: 
lobe-dominated quasars (LDQs); large solid circle: GPS source. 
The dotted horizontal line marks the division ($EW$ = 1\AA) between weak and strong absorbers
adopted in this work. The dot-dashed line shows the 95\% completeness level (0.5\AA).
\label{ew_l.fig}
} 
\end{center}
\end{figure*}

Figures~\ref{ewalfmv.fig} and~\ref{ew_l.fig} illustrate how the total 
absorption strength for each object distribute with the quasar properties: 
$\alpha_{\rm UV, \lambda}$ (Fig.~~\ref{ewalfmv.fig}a, c), $M_V$ 
(Fig.~\ref{ewalfmv.fig}b, d), and the 1550\AA{} monochromatic continuum 
luminosity, $L_{\rm cont}$(1550\AA) (Fig.~\ref{ew_l.fig}). Table~\ref{stats.tab}
lists the Spearman's correlation rank and the probability ($P$) that no correlation
is present.
The total strength (\ie restframe $EW$) of the \civ{} NALs in each quasar
in the combined sample of RLQs and RQQs correlates strongly with
$\alpha_{\rm UV}$ (Spearman's rank, r\,=\,0.53) with a $P$\,$<$0.1\% 
probability of occurring by chance (Fig.~\ref{ewalfmv.fig}a, c). 
Thus, more strongly absorbed objects tend to have redder continua.
The RLQs dominate this relationship (Table~\ref{stats.tab}). The
RQQs exhibit a weaker and less significant trend (see below); they 
clump\footnote{This may be due to our inability to select RQQs with a
range of source inclinations (\S~\ref{smpls}), since inclination
and absorption strength are likely related, as discussed in \S~\ref{inclin}.} 
around $\alpha_{\lambda} \eqsim -1.5$ (Fig.~\ref{ewalfmv.fig}a). 
Little difference is seen when individual absorbers with $EW < 0.5$\AA{} 
(the 95\% completeness limit) are omitted. These trends are consistent 
with a larger dust reddening in objects subject to stronger NALs. 
Notably, the RQQs with BALs or mini-BALs have some of the reddest UV
continua indicating they are subject to a larger dust reddening on
average.

The dependence of the total NAL strength ($EW_{\rm tot}$) for each object
on source
luminosity is different for the two quasar types (Fig.~\ref{ewalfmv.fig}b, d; 
Fig~\ref{ew_l.fig}); the two luminosity measures, $M_V$ and 
$L_{\rm cont}$(1550\AA), show consistent results. For RLQs the NALs 
tend to get weaker in more luminous objects
(Figs.~\ref{ewalfmv.fig}d and~\ref{ew_l.fig}b, 
Table~\ref{stats.tab}). The inverse $EW_{\rm tot} - L_{\rm cont}$(1550\AA) 
correlation is even more significant (r\,=\,$-$0.37 and P\,=\,0.9\%)
when the individual absorbers are tested (see also Paper~I).
However, the robustness of this luminosity dependence is unclear 
because (a) the relationship essentially disappears when 
the $EW_{\rm tot} <$\,0.5\AA{} absorbers are omitted (Fig.~\ref{ewalfmv.fig}d,
Fig~\ref{ew_l.fig}, right) and (b) it is not known with certainty that
the omitted absorbers (in item (a)) are associated with the quasar (\S~\ref{data}).
More importantly, there is a noticeable lack of $EW_{\rm tot} \geq 0.5$\AA{} 
absorption in quasars with $M_V \lsim -29$ mag.
This is in sharp contrast to the NALs in RQQs (Figs.~\ref{ewalfmv.fig}b 
and~\ref{ew_l.fig}a, Table~\ref{stats.tab}):
stronger NALs occur in more luminous RQQs. In particular, most of the
RQQs in Fig.~\ref{ewalfmv.fig}b with $M_V \leq -29$ mag have
$EW_{\rm tot} \gsim$ 3\AA. The $EW_{\rm tot} - $ luminosity correlation for
the RQQs is significant at the 95\% level (Table~\ref{stats.tab});
the NALs of the two RQQs with $EW_{\rm tot} < $ 0.5\AA{} are likely intervening.
This result is intriguing, if real, because an increasing absorber strength
is naively expected if increased radiation pressure in more luminous, bluer
objects is responsible for heating the accretion disk and evaporating 
more absorbing gas off the disk (\eg Paper~I; Laor \& Brandt 2002).
This is briefly discussed in \S~\ref{outflows}.


There appears to be an upper envelope in $EW_{\rm tot}$ for increasingly 
redder slope for the RLQs in Figure~\ref{ewalfmv.fig}c.
This suggests that strong NALs are avoided in very blue objects.
The individual subgroups of RLQs (CDQ, LDQ, CSS, GPS) are seen in 
Figure~\ref{ew_l.fig} to distribute fairly similarly within the covered 
values of luminosity and absorption strength, although the CDQs do not have
NALs with $EW \gsim$3\AA.

The distribution of RQQ NALs in Figure~\ref{ewalfmv.fig}a is consistent with
that defined by the RLQs in Figure~\ref{ewalfmv.fig}c, but the population of
weak absorbers in very blue RQQs ($\alpha_{\rm UV, \lambda} < -2.$), as 
seen for the RLQs, is missing. 
The single RQQ outlier at $\alpha_{\rm UV, \lambda} \approx -3$ has a 
low-velocity absorber which also absorbs part of the underlying continuum. 
This particular system may therefore be located outside the central engine, 
perhaps even in the quasar host galaxy, along our line of sight. This could 
explain why this absorber does not follow the trend outlined by the other, 
mostly intrinsic, absorbers.


\begin{figure*}[t]
\hspace{1.5cm}
\vbox{
\epsfxsize=13.0cm
\epsfbox{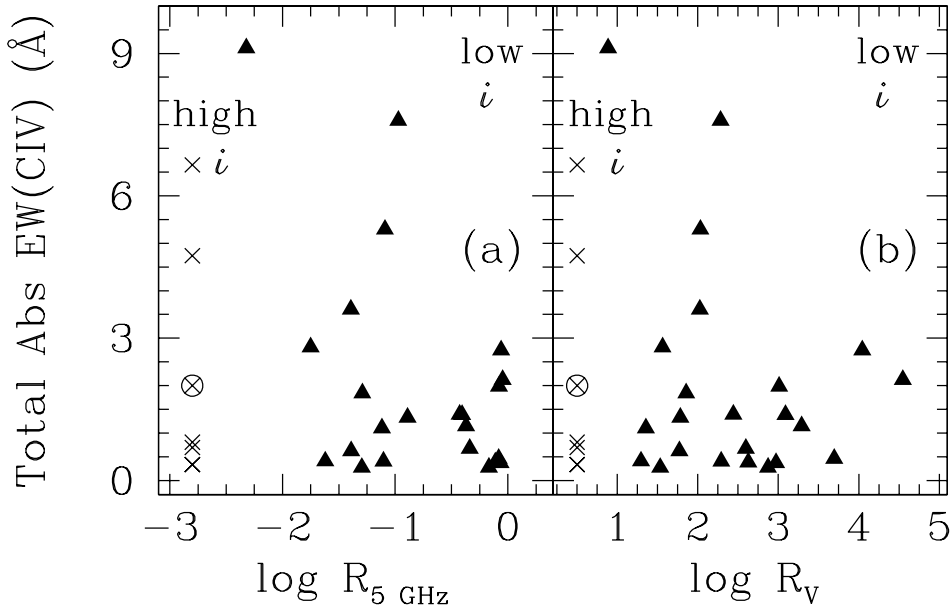} 
}
\begin{center}
\vspace{-0.5cm}
\caption[]{The total absorption strength for each RLQ versus estimators of the radio source 
inclination ($i$) relative to our line of sight, defined in \S~\ref{smpls}.  Only the subset 
of RLQs defined by 
Fig.~\ref{lext.fig} is shown. Recall, LDQs have $\log R_{\rm 5\,GHz} < -0.3$.
The absorption strengths of all the CSS objects in the full sample are shown as crosses 
at the very left part of each panel.  This is because the radio core luminosity is 
commonly not well determined for CSS quasars, inhibiting an estimate of $i$.  
Q1726$+$344 with a ``mini-BAL'' trough is encircled.
\label{lgr.fig}}
\end{center}
\end{figure*}

\subsection{Dependence on Radio Source Inclination \label{inclin}}


\begin{figure*}[]
\hspace{2.00cm}
\vbox{
\epsfxsize=12.0cm
\epsfbox{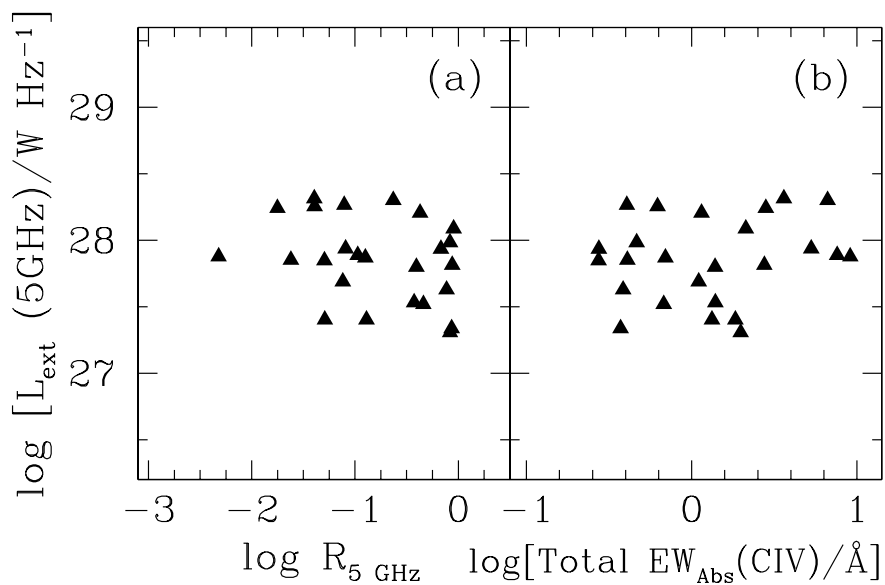} 
}
\begin{center}
\vspace{-0.5cm}
\caption[]{Extended radio luminosity shown against ({\it a}) the radio source
inclination estimator and ({\it b}) the total strength of the absorption
systems (for each quasar) for the subset of RLQs used in Fig.~\ref{lgr.fig} to test for
inclination dependence. There is no bias toward intrinsically brighter/fainter radio
sources with inclination angle (panel {\it a}). For a given intrinsic source brightness
there is a range of absorption strengths (panel {\it b}). 
\label{lext.fig}}
\end{center}
\end{figure*}

Figure~\ref{lgr.fig} shows how the total NAL strength in a subset of the RLQs
distribute with the two measures of radio source inclination described in
\S~\ref{smpls}.  
As discussed in \S~\ref{freq}, NALs are more frequently occurring in LDQs 
than CDQs.  Figure~\ref{lgr.fig} illustrates that LDQs also tend to be more 
strongly absorbed than CDQs.  This is not a new result (\eg Anderson \et 1987; 
Foltz \et 1988; Wills \et 1995; Barthel \et 1997; 
Ganguly \et 2001; cf.\ Baker \et 2002 and \S~\ref{radioevolution}). 
LDQs are believed to be intrinsically similar to CDQs, 
just viewed at a larger inclination, $i$, of the radio axis relative to 
our line of sight (\S~\ref{smpls}).  Hence, absorption strength may, in 
part, depend on source inclination.  Barthel \et (1997) used almost the 
same RLQ sample as that currently analyzed.  Here, the inclination dependence of the 
NAL $EW$s is tested on a more well-suited sub-sample of the RLQs than
done by Barthel et al. This is important if the NALs are intrinsic to the 
quasar central engine and, for example, depend on intrinsic source brightness.  
Therefore, to avoid possible selection biases the LDQs and CDQs 
were selected for this particular analysis to cover the same range in the 
luminosity of the extended radio emission, L$_{\rm ext}$, as illustrated in
Figure~\ref{lext.fig}a.
The extended emission, as opposed to the compact, nuclear ``core'' radio
emission, is isotropically emitted and is thus a better measure of the
intrinsic radio power.
Figure~\ref{lext.fig}b confirms that for this subset of RLQs 
the total NAL strength for each quasar is not correlated with L$_{\rm ext}$,
as expected for a sample unbiased in intrinsic radio power. Also, it
is comforting that there is no $EW$ dependence on the total radio luminosity, 
L$_{\rm radio}$(5\,GHz), both for the full sample of RLQs (see \eg Fig.~3 in Paper~I) 
and for the subset (not shown but the distribution is similar to that of the full 
sample).  Figure~\ref{lgr.fig} shows that even for this subsample, the 
strongest ($EW >$ 3\AA) NALs are seen in LDQs and the strength increases 
with\footnote{Fig.~4 in Paper~I shows this distribution for the strength of the 
individual absorbers.} inclination (lower $\log R_{\rm 5GHz}$ and $\log R_{\rm V}$); 
again, this is similar to the full RLQ sample (not shown but 
see \eg Fig.~1 by Barthel \et 1997). The fact that this trend exits in the subsample 
and the various non-restricted samples of NALs mentioned above shows that the trend 
is not due to intrinsically brighter or fainter lobe-dominated sources in those 
samples.  If indeed LDQs are relatively more inclined, then the correlations of NAL 
strength, $EW$, with $\alpha_{\rm UV}$, $\log R_{\rm 5GHz}$, and $\log R_{\rm V}$ 
(Figures~\ref{ewalfmv.fig} and~\ref{lgr.fig}) can be explained as a combination of 
inclination and reddening effects, which thus in RLQs seem to dominate possible 
radiation pressure effects on disk outflows; as argued in \S~\ref{absprops} a 
luminosity dependent absorption strength is expected in the latter scenario.


\begin{figure*}[t]
\hspace{2.00cm}
\vbox{
\epsfxsize=14.0cm
\epsfbox{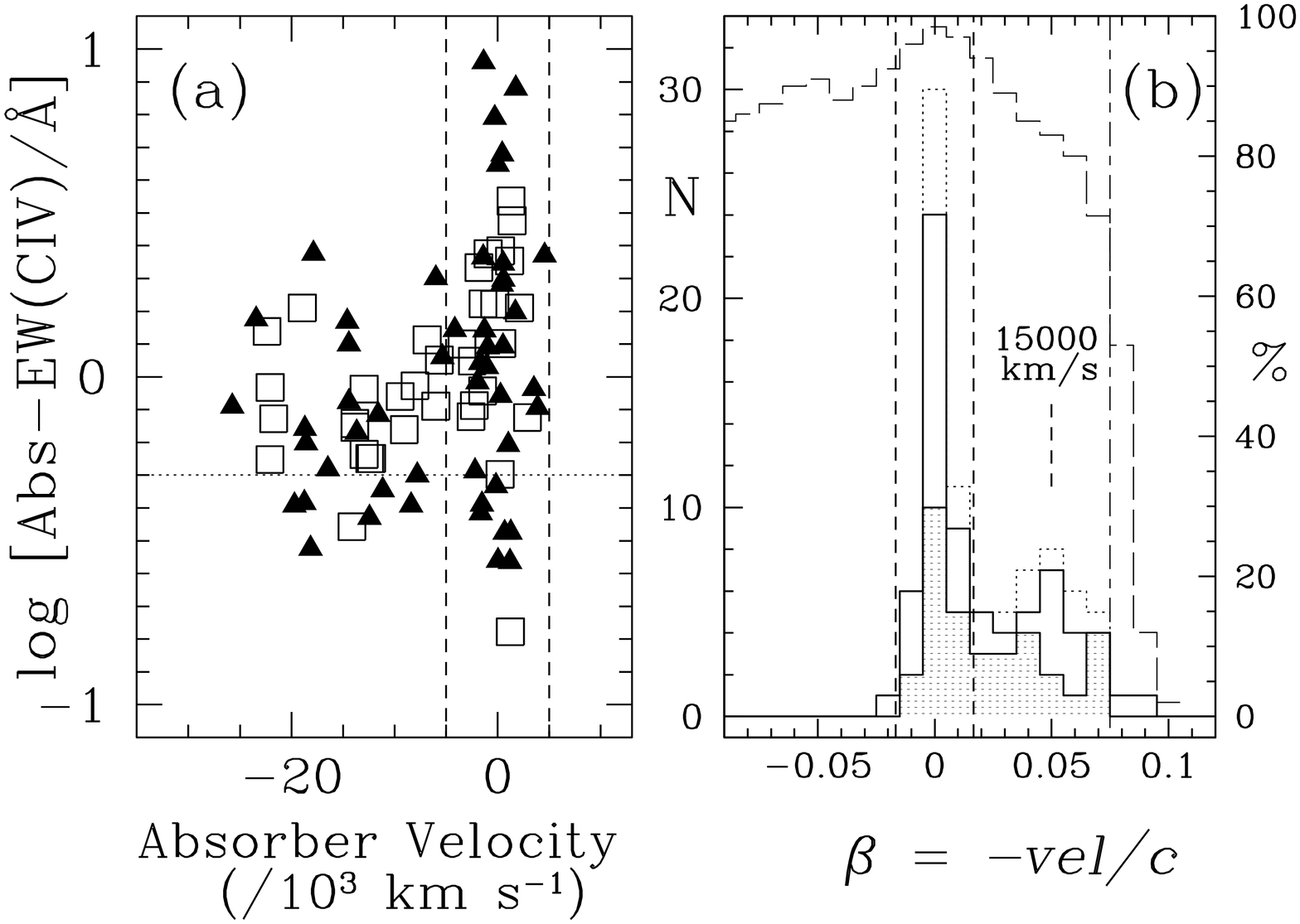}
}
\begin{center}
\vspace{-0.5cm}
\caption[]{(a) Relationship between the strength and velocity of the individual
\civ{} narrow absorber for RQQs (open squares) and RLQs (solid triangles). The velocity
space commonly considered for associated absorbers, $|vel| \leq 5000$ \kms, is marked.
The 95\% completeness level within $|\beta | \leq 0.02$ (6\,000\,km s$^{-1}$)
is $EW_{\rm rest} \geq 0.5$\AA{} (Fig.~\ref{ewdistrib.fig}). 
The horizontal dotted line marks this completeness level.
(b) Number distribution (left $y$-axis) of narrow absorber velocity,
parameterized as $\beta$ (defined in the text).
The dotted light style histogram represents the distribution with respect to $\beta$ 
for the full sample of absorbers among the RLQs and RQQs. 
The subsample of \civ{} doublet absorbers stronger than the completeness limit of 0.5\AA{} 
(rest-$EW$) are shown by the open solid histogram; the RQQ subset thereof is shown shaded.  
See Fig.~\ref{betarlq.fig} for the RLQ subsets.
The range in $\beta$ consistent with associated absorbers is marked by the two vertical 
dashed lines.  The dashed histogram (right $y$-axis) shows the completeness level of 
absorbers down to $EW_{\rm rest} = 0.5$\AA{} for a given $\beta$.  A completeness level 
greater than 70\% is obtained for $\beta \leq$ 0.07 (long/short dashed vertical line).
\label{vel.fig}}
\end{center}
\end{figure*}

Such related effects of absorption and reddening are not clearly present for the 
RQQ sample (Figs.~\ref{ewalfmv.fig} and~\ref{ew_l.fig}). However, the lack of, 
even crude, inclination measures for RQQs precludes a sample selection to test
inclination dependent absorption strengths. This likely explains the concentration
of $\alpha_{\lambda}$ in Figure~\ref{ewalfmv.fig}a. However, the BAL quasars
suggest that a similar inclination dependent distribution of reddening dust exists
in RQQs, since (a) the BALs are among the reddest RQQs (\S~\ref{absprops}), 
and (b) BAL outflows are likely equatorial (\S~\ref{intro}).
Furthermore, it is generally a worry that the optically selected quasars 
detected in early surveys are biased toward the brightest and hence the least 
dust reddened sources. This is because sources viewed almost face-on are 
likely more luminous and bluer, rendering them more easily detected when at 
large distances.  But Figure~\ref{ewalfmv.fig}a shows that the 
current RQQ sample tends to lack sources with very blue spectra 
($\alpha_{\rm UV, \lambda} < -2$) compared to the RLQs (Fig.~\ref{ewalfmv.fig}c). 
Also, while the most luminous and blue RLQs tend to have weaker NALs and a larger 
number of them (Figs.~\ref{ewalfmv.fig}c and~\ref{ewalfmv.fig}d), the 
RQQs of similar luminosity to the most luminous RLQs tend to have the 
strongest NALs (compare Figs.~\ref{ewalfmv.fig}b and~\ref{ewalfmv.fig}d),
as noted in \S~\ref{absprops}. 
Since BAL quasars tend to be relatively redder and less luminous (Brotherton \et 
2001; Goodrich 2001; Hall \et 2002), to have strong absorption, and are most
likely highly inclined (\S~\ref{intro}), could the 
omission of BAL $EW$s from Figure~\ref{ewalfmv.fig} explain the lack of a 
correlation similar to that seen for the RLQs?  This is not very likely as
BALs are {\it much} stronger than NALs. For example, the strength of the BAL
feature of the one BAL quasar that sneaked into the RQQ sample is $\sim$27 \AA{}.
Inclusion of this BAL in Figure~\ref{ewalfmv.fig}b would place it far from
the parameter space and trends defined by the NALs at no clear extension of the 
trend seen in Figure~\ref{ewalfmv.fig}a and~\ref{ewalfmv.fig}b.
However, the BAL strength and $M_V$ of this quasar is consistent with trends
defined by low-redshift BAL quasars (Laor \& Brandt 2002; Vestergaard 2004b).


\begin{figure*}[t]
\hspace{2.00cm}
\vbox{
\epsfxsize=14.0cm
\epsfbox{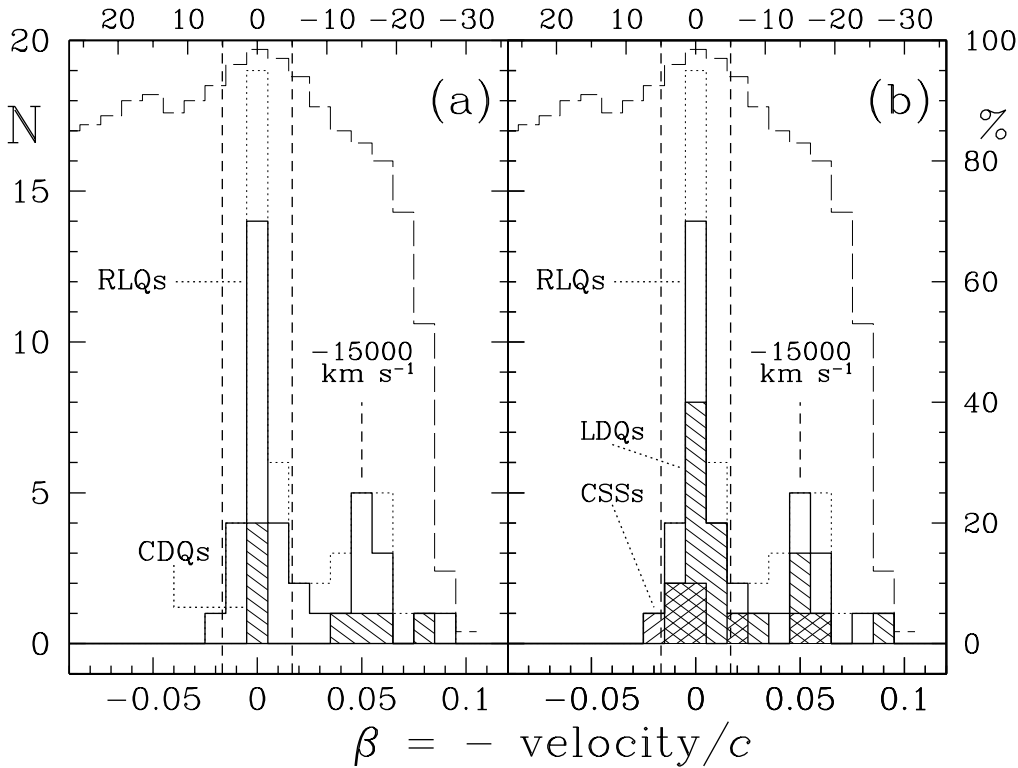}
}
\begin{center}
\vspace{-0.5cm}
\caption[]{Number distributions (left $y$-axis) of the $\beta$ parameter of the narrow absorbers 
detected in the RLQs and various radio quasar subsets, as marked. The absorber velocity is 
shown in the top $x$-axis in units of 1000 \kms{}. Only absorbers stronger than the 95\% 
completeness limit within $|\beta| \leq 0.02$  of 0.5\AA{} rest-$EW$ are included here
(solid histograms; dotted histogram: full sample of RLQ absorbers).
The dashed histogram (right $y$-axis) shows the completeness level of absorbers down to 
$EW_{\rm rest} = 0.5$\AA{} as a function of $\beta$.
\label{betarlq.fig}}
\end{center}
\end{figure*}

\subsection{Velocity Distribution of the Absorbers \label{veldistrib}}

In Figure~\ref{vel.fig}a the $EW$ of each NAL is shown as a function of the 
absorber velocity relative to the quasar \civ{} emission redshift.
The number distribution of individual $EW \geq$0.5\,\AA{} NALs with velocity
(parameterized by $\beta$) are shown for the various quasar types in
Figures~\ref{vel.fig}b and~\ref{betarlq.fig}.
The commonly adopted velocity limits ($\pm$5000\,\kms) for the associated 
absorbers (Foltz \et 1986) are marked.  The strongest NALs are clearly 
{\it associated} with the quasars (Fig.~\ref{vel.fig}a).  Set aside the 
five strongest NALs in RLQs, there is a similar velocity distribution 
for RLQs and RQQs.  The enhancement of associated NALs is evident and 
quantified in Figure~\ref{vel.fig}b.  A second peak is present at a 
velocity of $-$15\,000\,\kms{} (see also Fig.~\ref{vel.fig}a). The RQQs 
contribute significantly to this enhancement by having a large 
number of absorbers over a range of outflow velocities. The RLQs are 
characteristically strongly clustered around $-$15\,000\,\kms{} 
(Fig.~\ref{betarlq.fig}) with LDQs contributing the most. 
This clustering around $\beta \approx 0.05$ roughly coincides with 
the increase in absorber strength (Fig.~\ref{vel.fig}a) at a velocity 
around $-$18\,000\,\kms. 
Notably, this $EW$ increase is present for both RLQs and RQQs.

The long-dashed histogram in Figures~\ref{vel.fig}b and~\ref{betarlq.fig} 
shows the completeness level of NALs down to $EW_{\rm rest}$\,=\,0.5\AA{} for 
a given $\beta$ bin. 
The high completeness fraction ($>$\,70\%) across the entire second 
enhancement (both in $EW$ and $\beta$ parameter space) shows that
these enhancements are real. While there is a danger that the weaker NALs
($EW$\,$<$\,1\AA) at high velocity may be unidentified intervening systems, 
more so for the RQQs (\S~\ref{data}), there is little doubt that the stronger
($\gsim$ 1\,\AA), 
high-velocity absorbers are intrinsic to the quasars (see Figure~\ref{vel.fig}a).

\subsection{Are NALs Related to Quasar Outflows? \label{outflows}}

The more dramatic BALs are due to large column densities of outflowing, 
absorbing gas (\eg Hamann 2000). The quasar-intrinsic NALs may also 
be associated with
outflowing gas somehow, since (i) strong (\ie likely quasar intrinsic) 
high-velocity NALs exists, and (ii) the high-velocity enhancements at
$v \approx 15\,000 - 18\,000$\,\kms, discussed above, are intriguingly 
close to the typical BAL terminal\footnote{Outflow velocities up to 
$\sim$60\,000\,\kms{} have, however, been detected (\eg Jannuzi \et 1996; 
Hamann \et 1997).} outflow velocity ($\sim$20\,000\,\kms).
Also, the occurrence
and properties of NALs and BALs in RLQs and RQQs are generally consistent
with predictions of the disk-wind outflow model of Murray \& Chiang (1995).
Specifically, RQQs tend to have NALs of modest strength (\S~\ref{compltness};
Fig.~\ref{ewdistrib.fig}a; Fig~\ref{vel.fig}a) and, until recently, were
the only quasars known to have BAL systems.
Also, RLQs have quite strong associated NALs (\S~\ref{absprops}),
but rarely\footnote{Although Becker \et (2001) find $\sim$10\% BALs 
quasars among bona-fide RLQs (\ie the radio-loudness, 
$R^{\ast} = L_{\rm radio}$/$L_{\rm optical} >$ 10) almost similar to the
RQQ fraction ($\sim$10\% $-$ 15\%; \eg Hewett \& Foltz 2003),
most of the detected BAL quasars are radio-intermediate sources
($ 1 \leq R^{\ast} \leq 10$); the radio-intermediate nature appears common
among BAL quasars in general (Francis, Hooper, \& Impey 1993) often owing
to the absorption and reddening of the optical emission (\eg Goodrich 2001;
Becker \et 2001).  The relative distribution among radio type is unknown at
present, but since many are highly absorbed (see Hall \et 2002 for unusually
looking BAL quasar spectra from Sloan Digital Sky Survey), and apparently
radio-intermediate, most of the BAL quasars are presumably intrinsically
radio-quiet. } develop the strong BAL systems (\eg Goodrich 2001; but 
see Becker \et 2001).  
In the disk-wind model of Murray \& Chiang 
(1995), the stronger X-ray flux in RLQs reduces the radiation 
pressure on the disk winds by stripping the atoms off electrons. In effect, 
RLQs are incapable of accelerating the high-density outflows to relativistic
velocities and strong, low-velocity absorption is predominantly expected
instead. This provides a reasonable explanation for the relative
strengths of the associated, narrow absorbers among RLQs and RQQs seen here.
In this case, there is an immediate connection between NALs and BALs
as the RLQ associated NALs are ``failed'' BAL systems. (Other, related
scenarios were suggested by Elvis 2000; Ganguly \et 2001; Laor \& Brandt 2002).

The moderately strong, high-velocity absorbers seen at velocities
close to the BAL terminal velocities    
add an interesting twist to
this interpretation. Since these high-velocity NALs occur in
both RLQs and RQQs, they may originate in a second, perhaps different, 
population of narrow absorbers common to quasars irrespective of radio
type. Figure~4 in Paper~I shows that these high-velocity absorbers are not 
confined to objects at a specific source inclination; Figure~\ref{betarlq.fig} 
also shows a mix of CDQs and LDQs with high-velocity NALs.

To explain the presence of high-velocity (associated) NALs in both RLQs and RQQs, 
two scenarios appear possible: entrainment by BAL outflows or by radio 
outflows, as discussed in turn next. 
First, while one may imagine the high-velocity RQQ NALs, typically with 
$EW \lsim$3\AA, originate in discrete, turbulent clouds of relatively lower 
column density gas entrained by the BAL outflow, this cannot easily
explain the high-velocity absorbers in RLQs: BAL outflows are rarer 
and some shielding from the fatal X-ray flux would seem necessary 
to allow radiation pressure acceleration of these NAL systems.
Besides, BAL outflows are {\it not} detected in these RLQs with 
high-velocity NALs.  Alternatively, these high-velocity
absorbers may originate in gas stirred up and perhaps entrained by the
radio jet. However, such a scenario appears a little contrived given the
similar occurrence in RQQs, unless these RQQs in fact have
some weak radio source outflow which could possibly entrain NAL gas.
To test this, the NED\footnote{NASA/IPAC Extragalactic Database} radio 
data base was searched for indications that the two RQQs with strong 
($EW >$ 1\AA{}) high-velocity NALs (Figure~\ref{vel.fig}a) are radio sources. 
Neither of the FIRST nor NRAO VLA Sky Survey 1.4\,GHz radio images reveal any 
radio emission down to 1\,mJy within $\sim$1 arcminute of the optical position 
of the RQQs.
Deeper radio imaging at lower frequency is needed to determine whether 
these RQQs may have some weak, uncollimated radio outflows which could 
possibly account for accelerating the intrinsic NAL absorbers.
Instead, Ganguly \et (2001) propose that RQQs have higher mass-loss rate, a 
higher wind velocity, and in effect have winds with higher matter
densities compared to RLQs. While this may explain the dearth of
BALs in RLQs, it does not explain the intermediate and weak NALs
($EW_{\rm rest} \lsim$ 3\AA) at high outflow velocities seen for
both RQQs and RLQs. Also, Vestergaard (2004b) finds no support
for RQQs typically having higher \mdotratio{} than RLQs in this 
$z \approx 2$ quasar sample. 

In summary, the presence of strong associated NALs in RLQs and the 
presence of relatively weaker associated NALs in RQQs are quantitatively
consistent with expectations of 
radiatively driven disk-wind scenarios. The $\sim$1\AA{} high-velocity
NALs may originate in a second population of narrow absorbers. 
Entrainment by either BAL outflows or radio plasma outflow do not 
appear to uniquely and easily explain their presence in both RLQs 
and RQQs.

\section{Is Absorption Related to Radio Source Evolution? 
\label{radioevolution}}

One subset of the RLQs, the CSSs, may be able to shed light on whether
absorption in RLQs is related to the onset and propagation of the radio
source (Baker \et 2002). The reason is that CSSs are possibly\footnote{
The alternative interpretation is that a very dense intergalactic 
medium confines the radio source to a very small physical size, preventing 
significant radio source growth (``frustrated radio source''). However,
the data currently appear to favor the young source interpretation (see 
O'Dea 1998 for a review).}
young (O'Dea 1998) and, as can be
seen in Figure~\ref{lgr.fig}, CSS absorbers exhibit
a range of strengths and some are relatively strong.
Baker \et (2002) propose that associated NALs in CSSs are in fact related to 
the radio source evolution, based on their study of \civ{} $\lambda$\,1549 
associated absorbers for RLQs at $0.7 < z < 1.0$ and $1.5 < z < 3.0$ 
detected in the well-defined Molonglo radio survey (Kapahi \et 1998). 
The original low-frequency (408\,MHz) selection and high completeness level
($\sim$97\%) down to the survey limit of 0.95\,Jy in a well-defined area on
the sky make the Molonglo radio survey one of the least biased surveys with 
respect to radio source inclination, $i$ (Baker \et 2002). 
Baker \et find that the most strongly absorbed objects ($\lsim 5$\AA{})
in their sample are both the most dust reddened and have the smallest radio 
sources, most of which are CSS in nature. While Baker \et do find their LDQs 
to be more strongly absorbed than
their CDQs (similar to earlier findings and this study; see \S~\ref{inclin}), 
the CSS quasars are
after all the most strongly absorbed. The authors propose to see an effect of
radio source evolution in which the young radio source is enshrouded in gas and
dust: at early radio source age the nuclear spectrum is seen through a large
amount of reddening dust and absorbing gas. But with time, as the radio source 
grows, this cocoon is shed and dispersed or destroyed leaving less material
(or covering fraction thereof) to absorb the centrally emitted spectrum.

\begin{figure*}[t]
\hspace{1.5cm}
\vbox{
\epsfxsize=14.0cm
\epsfbox{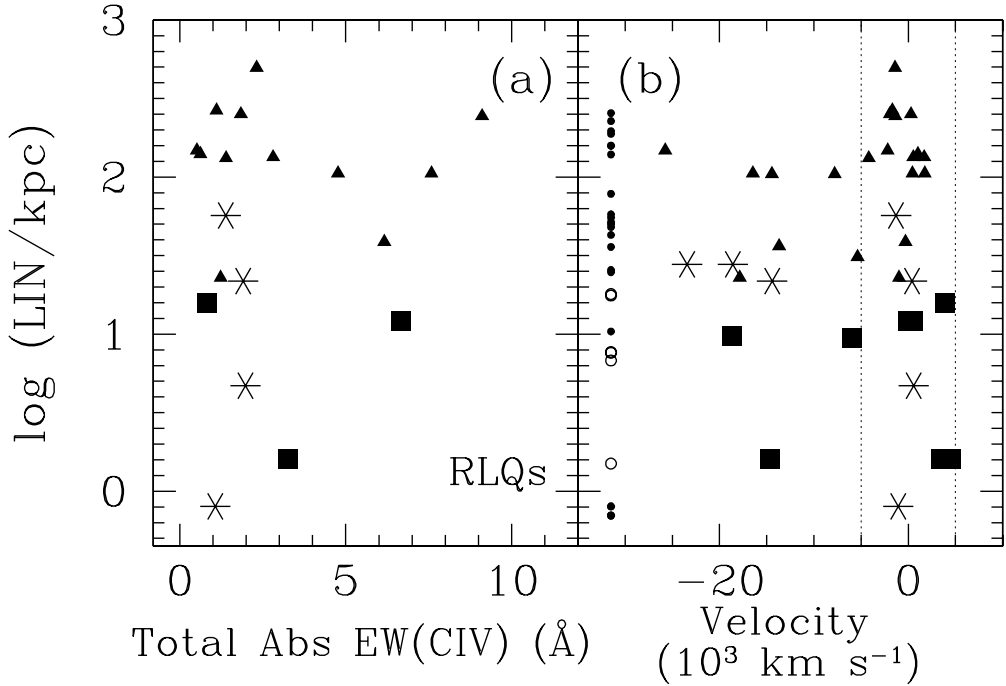}
}
\begin{center}
\vspace{-0.5cm}
\caption[]{Linear radio source size shown relative to ({\it a }) total absorption strength
for each RLQ and relative to ({\it b}) the individual absorber velocity.
Only absorbers stronger than the 95\% completeness limit within $|\beta| \leq 0.02$  
of 0.5\AA{} rest-$EW$ are included here (both panels). In addition, only {\it associated}
absorbers are included in panel (a) to ease a direct comparison with Baker \et (2002).
(The absorbers omitted from that panel mostly occupy the region below LIN of 100\,kpc and
$EW \lsim 2$\AA.) Note that the use of the total $EW$ for each quasar is appropriate
for assessing which quasars are more strongly absorbed and is fully
comparable to the Baker \et study.
The distribution of linear sizes of unabsorbed RLQs (open circles: CSS;
solid circles: other RLQs) are shown at a relative velocity of $-$30\,000 \kms{} for reference. 
Symbols: solid squares: CSS; solid triangles: LDQs; asterisks: CDQs. 
Note: to aid a direct comparison with Baker \et (2002), $H_0$ = 50 ${\rm km~ s^{-1} Mpc^{-1}}$, 
q$_0$ = 0.5, and $\Lambda$ = 0 is adopted for the linear sizes, which are $\simeq 0.61 \times$
LIN($H_0$=50;q$_0$=0.0).
\label{linsize.fig}}
\end{center}
\end{figure*}

Curiously, the inverse trend between linear radio source size and absorber
strength seen by Baker \et{} is in sharp contrast to the data studied here.
Figure~\ref{linsize.fig} shows how the total associated-absorption strength 
for each RLQ
distributes with the `largest linear radio source size' (Miley 1971); the
completely opposite trend is evident.  The CSS quasars show a range 
of absorption strengths, some of which in fact have high outflow velocities
(Figure~\ref{linsize.fig}b).  The GPS and CSS phenomena are possibly related 
(\eg O'Dea 1998) and the one absorbed GPS object also has a high
velocity, but relatively weak, absorber.  Although two of the CSSs have similar 
or stronger absorption than the CSS objects in the Baker \et (2002) sample, the 
LDQs with large radio sources after all have the strongest 
associated NALs.  In fact, there appears to be a lower envelope in the linear 
size for a given absorber $EW$ for the current sample.  
Furthermore, there is no apparent relation between absorber velocity
and linear radio size for this sample (Figure~\ref{linsize.fig}b).


In an attempt to better understand the discrepant results with the Baker \et
study, the properties of the two RLQ samples have been compared; the focus was 
placed on the high redshift ($1.5 \leq z \leq 3.0$) subset of the Baker \et quasars
which has similar redshift range to the current RLQ sample.
An important issue is that of sample selection.
Most of the RLQs studied here were originally detected in low-frequency 
(178\,MHz and 408\,MHz) surveys (\S~\ref{data}) similar to the Molonglo quasars. 
Also, while the RLQ sample studied here is non-homogeneous and incomplete,
this nature is both a liability and an asset. The sample does have the 
advantage that it is selected to reduce selection biases. For example: (a)
the most beamed sources are not included, (b) RLQs with a range of source 
inclinations were selected (\S~\ref{data}), and (c) some control over the 
intrinsic brightness is employed (\S~\ref{data} and \S~\ref{inclin}). However,
these RLQs are also expected to be more luminous than the Molonglo sources,
owing in part to their original detection in earlier surveys that were less 
sensitive than more recent surveys commonly are, and this also appears to be 
the strongest difference between the two samples.  A comparison based on $M_V$ 
and 408\,MHz radio luminosities, 
$L_{\nu}$(408MHz), confirms this (Table~\ref{qsamples.tab}): this RLQ sample
is more luminous on average by $\sim$1\,dex in $L_{\nu}$(408MHz) with very 
little overlap with the Baker \et values, and by 1.5\,magnitude in $M_V$. 
In summary, the Molonglo quasars are intrinsically fainter (radio and
optical) sources. Also, they have a higher fraction of CSS quasars (11/20) 
compared to this RLQ sample (14/66; Table~\ref{freq.tab}). Hence, the 
Molonglo sample contains a larger fraction of young sources, if the  
Baker \et scenario holds.

Assuming low-frequency selected, weaker radio sources are typically 
younger\footnote{
How may low-frequency surveys be more sensitive to relatively younger radio 
sources? Presumably, a young source with poorly collimated radio outflow will 
emit predominantly isotropic, steep spectrum radio emission.  We may be seeing 
this already for the (very faint) radio-sources in RQQs (\eg Barvainis, 
Lonsdale, \& Antonucci 1995; Kukula \et 1998). Likewise, a source in which the 
infant radio jet is so small that it contributes negligibly to the overall radio 
power would also tend to emit mostly isotropic lobe emission. In both cases 
these steep spectrum sources are most easily detected in low-frequency surveys. 
}, 
can radio source age explain the different absorption characteristics seen in 
these two studies, such that the current RLQs are relatively older? 
In that case, the Molonglo quasars should (1) be more frequently absorbed,
(2) be more strongly absorbed, and (3) be dust reddened to a higher degree.
The  Baker \et  $1.5 \leq z \leq 3.0$ sample not only has a higher fraction
of CSS quasars with associated NALs (8/11 $\simeq$73\% $\pm$25\%), but the
quasars are also more frequently absorbed in general (18/20 $\simeq 90$\%$\pm$21\%);
in comparison, only 36\% $\pm$ 16\% of the CSSs and 39\% $\pm$ 8\% of the RLQs
here have associated NALs, even when counting all the detected NALs (these 
fractions are not shown in Table~\ref{freq.tab}) as opposed to the most complete
subset ($EW \geq 0.5$\AA{}; \S~\ref{compltness}). However, the occurrence of
CSS NALs are marginally consistent between the samples to within the (large)
counting errors.
On the strength of the NALs, the Molonglo quasars are generally {\it less}
strongly absorbed: none of the Baker \et non-CSS quasars display absorption 
stronger than 2\AA, while many of the LDQs and CDQs studied here are more strongly 
absorbed. However, the NAL strength of the CSS quasars is comparable between 
the two samples (\eg compare Figure~\ref{lgr.fig} [$EW \lsim 7$\AA] with 
Figure~7 by Baker \et [$EW \lsim 5$\AA]). 
Regarding the degree of dust reddening, as judged from the UV continuum slope 
($\alpha_{\lambda}$) the Molonglo quasars are not subject to a significantly
higher dust column than the current RLQs: the range of 
$\alpha_{\lambda}$ measured among the absorbed RLQs is relatively similar 
between the two studies. Specifically, the absorbed Molonglo quasars cover the 
approximate range $-2 < \alpha_{\lambda} < 1$, which overlaps with most 
of the spectral slopes measured here for the RLQs (Figure~\ref{ewalfmv.fig}; 
note that also the CSSs display a range of slopes).
However, the trend between reddening and absorption strength in the Baker 
\et sample is fully consistent with the data presented here.
Since the RLQs in this study also tend to be more 
strongly absorbed, radio source evolution do not appear to explain the 
differences seen with the Baker \et study. 
This is supported by the model-predicted (Begelman 1999) and 
observed (\eg O'Dea 1998) negative luminosity evolution with radio source 
growth.  Hence, other effects must be at play and dominate possible radio 
source age differences between the samples.

An alternative possibility is that the deviating results are due to a 
statistical fluke in either or both studies as the results are after all 
based on small number 
statistics: in regard to NALs with $EW >$3\AA{} the Baker \et sample has 
six NALs of which two are detected in $z > 1.5$ CSSs with $EW$s larger than 
that of a single $z > 1.5$ LDQ narrow associated absorber with $EW > 3$\AA{} (Figure~8   
by Baker et al.). In comparison, this RLQ sample has six NALs, two of which 
are observed in LDQs with $EW$s larger than that of the two CSS NALs 
(Figure~\ref{linsize.fig}). 

In summary, both RLQ samples exhibit stronger absorption in more dust 
reddened objects, but the current sample does not support the findings
of Baker \et that CSS quasars are the most strongly absorbed sources.
Also, it does not appear likely that radio source evolution explains
the different absorption properties between the samples as expected if the
RLQs studied here are the more aged cousins
of the presumably young radio sources detected in the low-frequency
Molonglo survey. Rejecting the possibility that the differences are 
due to a statistical fluke, one is left to conclude that luminosity 
differences and sample selection most likely are the leading causes of 
the differing results obtained. The reason is that the RLQs in this work
are on average more luminous in the radio by about a factor 10 and in 
the optical by about a factor 3 (Table~\ref{qsamples.tab}) than the 
Baker \et sample.  The higher luminosities may somehow make the current 
RLQs more capable of producing stronger absorption. This issue is 
addressed by Vestergaard (2004b).

\subsection{Speculations}
\subsubsection{Luminosity Driven Absorption?}

How may a higher source luminosity of the RLQs in this study explain their
stronger absorption relative to the Molonglo quasars?
More luminous sources may subject the 
accretion disk and the immediate environment of the central engine 
(including a `torus', if present) to a larger radiation pressure which could 
generate and accelerate larger amounts of absorbing material (\eg Proga \et
2000; Laor \& Brandt 2002). While the properties of the absorbed RLQs in 
this study seem to show that radiation pressure effects do not dominate
over inclination dependent distributions of reddening and absorbing material, 
the RQQ NALs after all do support the radiation pressure interpretation 
(\S~\ref{absprops}, \S~\ref{inclin}). 
Hence, one may speculate that the luminosity dependent absorption trend 
(allegedly caused by radiation pressure-driven, absorbing disk outflows) is 
only visible for RLQs across several orders of magnitude in luminosity, as 
opposed to the relatively narrow ranges ($\lsim$1\,dex) studied here and by 
Baker \et (2002). 
Alternatively, the strong, associated NALs detected in the more luminous, 
larger scale radio quasars may be of a different origin than the CSS absorbers. 

However, an investigation based on well-selected quasar samples specifically 
targeted to address these issues are required to disentangle the effects of
radio source age from effects owing to the mass, luminosity, size,
and/or inclination of the source, in addition to the complicating effects of 
merger history and star formation on the amounts of matter available 
for reddening and absorption. It is hoped that these intriguing differences 
discussed above will motivate further studies of these issues, because 
an understanding of the evolution of active nuclei and quasars is important
for our understanding of many other phenomena, including star formation
and galaxy evolution. 

\subsubsection{Are Young CSS and BAL Quasars Connected?}

On a slightly different but relevant note, it is interesting
to speculate that the recently discovered high-ionization BAL quasars
among the 1.4\,GHz selected FIRST sources (Becker \et 2001) are in fact 
CSS quasars, even if most are radio-intermediate and not bona-fide RLQs. 
The reason is that BAL quasars display compact, often unresolved, radio 
structure, yet they essentially all have steep radio spectra. In addition,
for low-ionization BALs there is evidence that the BAL outflow phenomenon 
is connected with a recent merger and/or starburst in the host galaxy 
(Canalizo \& Stockton 2002), which may also trigger a radio source 
(Baker \et 2002). 
While this is much harder to establish for the more distant
high-ionization BALs, it is indeed an intriguing possibility. The
radio-loud, high-ionization BAL quasar, FIRST J1556$+$3517, may be 
the first known example (Najita \et 2000).
This interpretation is, however, not immediately consistent with the
lack of BAL quasars in the Molonglo survey, unless the RLQ BAL phase 
is very short.


\begin{figure*}[t]
\hspace{1.5cm}
\vbox{
\epsfxsize=14.0cm
\epsfbox{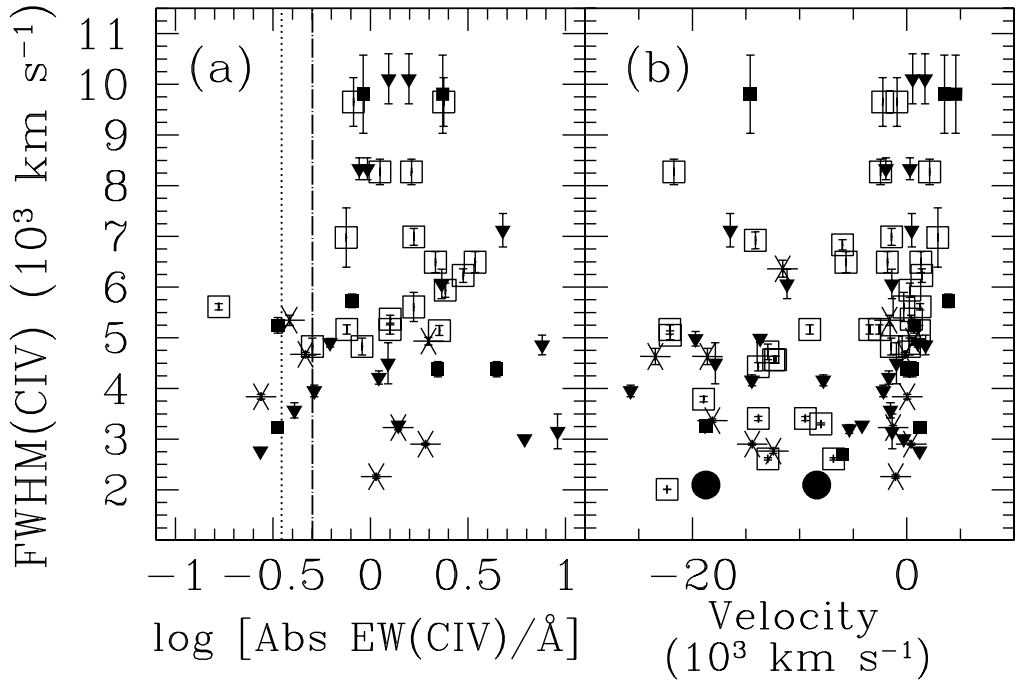} 
}
\begin{center}
\vspace{-0.5cm}
\caption[]{Distribution of \civ{} ({\it a}) individual narrow {\it associated} absorption strengths (i.e.,
$|v| \leq 5000$\,km s$^{-1}$) and ({\it b}) individual absorber velocity ( = $- \beta c$) with the 
\civ{} emission line width of the quasars.
The lower detection limit of 0.35\AA{} for CDQs with FWHM(\civ) $<$ 6000\,\kms{}, as determined
by Ganguly \et (2001), is marked by the dotted vertical line. The detection limit of 0.5\AA{}
as determined for the current quasar sample is marked with the dot-dashed vertical line.
Symbols: open squares: RQQs; solid triangles: LDQs; asterisks: CDQs; small solid squares:
CSS; solid large circles: GPS.
\label{fw_ewvel.fig}}
\end{center}
\end{figure*}

\subsubsection{On the Possible NAL Connection with Radio Source Evolution 
\label{nalorigin}}

If the onset of nuclear activity simultaneously triggers the radio source
and an initial BAL phase as suggested above, then at least the dynamical 
time-scales for these outflows should be consistent.  For the radio-quasar 
3C\,191 (Q0802$+$103), with a strong ($\sim$7\AA) associated NAL in its 
spectrum, the available data allow a crude test thereof. 
Indeed, dynamical time-scale estimates suggest that 3C\,191 could 
have started its nuclear activity expelling a BAL wind (or a birth cocoon) 
that is now observed as a distant, slow-moving NAL wind.
Specifically, this requires 
(a) that the low-velocity ($v \lsim$1400\,\kms) absorbing gas started out 
as a high-velocity BAL outflow ($v \gsim$20\,000\,\kms) that later 
decelerated significantly before reaching its current location 
$\sim$28\,kpc from the central engine (Hamann \et 2001), 
(b) that the radio source expanded\footnote{  
Radio lobe advancement speeds are never larger than 0.1$c$ 
(Longair \& Riley 1979).  Hot spot advancement speeds for compact
steep spectrum sources are about and not much higher than $\sim 0.02 c$
(Readhead \et 1996). Since these compact, likely young, sources are not
located in an environment particularly different from other radio sources
(\eg O'Dea 1998), these expansion speeds should be representative.
For example, the advancement velocity of the radio lobes in Cygnus\,A is 
comparable (Readhead \et 1996).}  
with a constant speed of $\sim$0.02$c$ (Readhead \et 1996) to its current 
linear size of $\sim$40\,kpc (Lonsdale \et 1993), and 
(c) that the optical-to-X-ray slope, $\alpha_{\rm OX}$, was much steeper 
at early epochs, such that this slope flattened as the quasar evolved. 
This is because BAL quasars typically have steeper slopes 
$\alpha_{\rm OX} \gsim 2$ (Brandt \et 2000; Green \et 2001) than that
observed for 3C\,191 ($\sim$1.4; Wilkes \et 1994). Also, flat 
$\alpha_{\rm OX}$ are not conducive to generating high-velocity BAL outflows 
according to the radiatively driven disk-wind model of Murray \& Chiang (1995). 
In addition, Hamann \et (2001) argue that the outflow must have swept up some
galactic material, since its current mass exceeds that expected for BAL outflows.  
Interestingly, these authors find their estimates of the size, mass, and 
velocities of the NAL gas in 3C\,191 to be consistent with the inferred values 
of these quantities for galactic superwinds in low-$z$ starburst galaxies.
Alternative scenarios to explain the 3C\,191 absorber are discussed by Hamann \et (2001).

It is currently unknown whether the steep $\alpha_{\rm OX}$ in BAL quasars
(a) is an intrinsic property of quasars capable of generating BAL outflows,
as predicted by disk-wind models, or (b) is caused by the absorbing mass 
outflow itself (\eg Brandt \et 2000). In the latter case, the nature of the 
outflow will need to change with time to explain the flat $\alpha_{\rm OX}$
currently observed for the above scenario to be viable.

\section{Emission Line Width Dependent Absorption? 
\label{fwhm}}

Ganguly \et (2001) find a significant lack of associated absorbers among 
CDQs with \civ{} emission line width (FWHM(\civ)) less than 6000\,\kms{} 
for the 15 quasars with \civ{} NALs from their \hst -observed sample of 
59 sources at $z \leq$1.  This observation can be explained if the 
gaseous region, responsible for the associated NALs, is located away 
from the radio jet axis (see Ganguly \et 2001 for details). In that 
case, our line of sight will not intercept the region occupied by the 
NAL systems for these sources, since CDQs are viewed at relatively 
smaller angles to the radio jets (\S~\ref{smpls}). 
It is important for our understanding of the central engine and its
evolution to know if such a trend is seen for more distant quasars as
well.  Table~\ref{fwfreq1.tab} lists for different subsamples the 
number of absorbed quasars with FWHM(\civ) smaller or larger than 
6000\,\kms{}, respectively. These subgroups are further subdivided 
based on the absorber velocity.  For reference, Table~\ref{fwfreq2.tab} 
gives the distribution of narrow-lined and broad-lined quasars 
among the absorbed and non-absorbed quasars. Note that in both tables
only quasars with NALs stronger than 0.5\AA{} are counted, since this 
is the most complete subset of NALs. Figure~\ref{fw_ewvel.fig} shows
how the strength and the velocity of the associated NALs distribute with 
FWHM(\civ) of the quasars.

Table~\ref{fwfreq1.tab} shows the striking result that for all subsamples
$-$ CDQs included $-$ and for all absorber velocities the quasars with 
relatively ``narrow'' \civ{} emission lines (cols. 4 and 7) are {\it more} 
frequently absorbed than broader-lined quasars (cols. 3 and 6), in stark 
contrast to the results of Ganguly et al. 
In fact, the most strongly absorbed quasars have FWHM(\civ) 
$\lsim$5000\,\kms{} (Figure~\ref{fw_ewvel.fig}a), and all CDQs with 
associated NALs ($EW \geq$0.5\AA) are ``narrow-lined''. 
After all, the broad-lined and narrow-lined CDQs are equally frequently
absorbed ($\sim$33\%; Table~\ref{fwfreq2.tab}, cols. 6 and 7).
But when all detected NALs are taken into account, as many as $\sim$2/3 of the
narrow-lined CDQs are absorbed. In addition, Table~\ref{fwfreq2.tab} shows
that with a lower $EW$ cutoff at 0.5\AA{}: (a) narrow-lined quasars are
more frequently {\it unabsorbed} among all the subsamples (cf.\ col.~4 and
col.~6), (b) the RQQs
display an even cut ($\sim$40\% within the errors) of absorbed objects with low and
high FWHM(\civ) (cols. 6 and 7), and (c) broad-lined RLQs as a group are more frequently
absorbed, owing mostly to the fact that all the broad-lined LDQs and CSSs 
are absorbed; however, care should be taken as the broad-lined RLQ subgroups 
have large uncertainties. 
Also, objects with different FWHM(\civ) exhibit no significant differences 
in absorption strength (Figure~\ref{fw_ewvel.fig}) with the exception that 
objects with very broad emission lines lack weak absorbers, 
$EW \lsim 0.6$\AA{} (Figure~\ref{fw_ewvel.fig}a). The CDQs tend to have 
FWHM(\civ) $<$ 7000\,\kms{} but also have NALs with $EW$ up to $\sim$2\AA.

In conclusion, the quasar sample here does not support the result of 
Ganguly \et (2001) that associated NALs avoid narrow-lined CDQs,
although narrow-lined quasars are slightly less absorbed, in general. 
Specifically, the narrow-lined CDQs with associated NALs are more 
frequently absorbed than the broad-lined CDQs, a property that all
the quasar subtypes share. This conclusion does not change if all
detected absorbers are analyzed.

The obvious possible explanations for the different results of our 
studies are: (a) a luminosity effect in the Ganguly \et sample causing 
an absorber cutoff for sufficiently luminous quasars, similar to that 
seen in Figure~\ref{ewalfmv.fig}d, and (b) different sample selections.
Although there is an indication that NALs stronger than $EW \gsim 0.5$\AA{}
are avoided in the most luminous ($M_V < -$27 mag) Ganguly \et quasars, 
it is unclear, and at most tentative, whether or not the luminosity 
effect mentioned in (a) is present. The reason is that the Ganguly
\et NALs are few and exhibit a large distribution in the $EW - M_V$
plane\footnote{$M_V$ was computed based on the data presented by
Ganguly \et (2001).} with no distinct correlation.  More importantly, 
the three absorbed flat-spectrum quasars ($\simeq$ CDQs) do not have 
distinct $M_V$ values from their unabsorbed siblings.  In terms of sample 
selection, the current sample is slightly more optically luminous than 
the Ganguly \et quasars (Table~\ref{qsamples.tab}), which may suggest
that low-$z$ quasars are subject to lower amounts and/or different
distributions of absorbing gas compared to the distant and more
luminous quasars, as also noted by Ganguly \et (2001).
However, sample selection is likely a significant factor considering
the fact that the low-redshift BQS quasars display a high absorption
frequency and have NAL strengths (Laor \& Brandt 2002) similar to
those reported here.

\section{Summary and Conclusions \label{conclusion}}

Weak and strong absorption is relatively common among low-redshift
active galaxies, but the fraction of quasars with associated
narrow-line absorption is still
poorly known, especially at moderate to high redshift.
The basic statistics of narrow \civ{} \lam 1549 absorption lines
detected in moderate redshift ($1.5 \lsim z \lsim 3.5$) radio-loud
and radio-quiet quasars are presented here (Tables~\ref{freq.tab}
and~\ref{freq3.tab}). This sample is
particularly suitable for this study for several reasons:
(1) it was {\it not} selected on account of the presence of
absorption in the restframe UV spectra, (2) the radio-loud quasars
have a large range in radio source inclinations and a significant
subset thereof can be defined where intrinsic brightness does
not bias inclination dependency tests, and (3) the radio-loud
and radio-quiet quasars were selected to match well in redshift
and luminosity, minimizing biases in the study due to these two
parameters and allowing a fair comparison of the absorption
properties of the two radio types.
It is important to note that this sample is not complete which
may affect the absolute absorption frequencies. However, given
the sample selection and the current lack of such statistics,
the absorption frequencies should provide fair guidelines until
complete samples are analyzed.
The main conclusions can be summarized as follows:

(1)  
Moderate redshift quasars display a high frequency ($\gsim$ 52\% $\pm$7\%) of 
narrow \civ \lam 1549 absorption lines (NALs), similar to Seyfert~1 
galaxies at low redshift and slightly higher than currently observed for 
the Bright Quasar Survey (BQS) quasars at $z \lsim$ 0.5.  The strength 
of the narrow absorbers is comparable to those detected in the BQS sample.
When restricted to the most complete subset of absorbers ($EW \geq$0.5\AA; 
completeness level $\gsim$95\%) about 40\% ($\pm \sim$9\%) of the quasars 
are absorbed, irrespective of radio type (\ie including radio-quiet, 
radio-loud, radio core-dominated, radio lobe-dominated, and compact 
steep-spectrum sources).  

(2) 
Strong narrow absorbers with $EW_{\rm rest} \gsim$ 1\,\AA{} are detected
in $\sim$35\% ($\pm$5\%) of the quasars with a slightly lower frequency 
among radio-quiet quasars ($\sim$30\% $\pm$8\%) compared to radio-loud 
quasars ($\sim$40\% $\pm$8\%).
However, a constant $\sim$25\% ($\pm$5\%) of the quasars, irrespective of 
radio type have strong {\it associated} absorption 
(velocity, $|v| \lsim$ 5000\,\kms) within the uncertainties (Table~\ref{freq.tab}).
Similar statistics are obtained for the $EW \geq$0.5\AA{} narrow absorbers.

(3) 
More strongly absorbed quasars tend to have redder UV continua,
confirming earlier findings based on radio sources only (Baker \et 2002)
and consistent with trends seen for broad absorption line quasars
(Brotherton \et 2001).

(4)
Lobe-dominated quasars, believed to be viewed at relatively larger angles 
to their radio source axis than core-dominated sources, are most frequently 
and most strongly absorbed in general. While this has been seen previously, 
the trend was tested here
on a well selected subset of radio-loud quasars with a range of source
inclinations and with luminosity biases strongly reduced. This solidifies
the reality of this trend, indicating a near-equatorial distribution of
narrow absorbers.

(5) 
Among the absorbed radio quasars significant correlations are found between
the strength of the absorbers and (a) the UV continuum luminosity (when all 
detected NALs are analyzed), and (b) the UV spectral slope, such that absorbed 
radio quasars tend to have redder 
and fainter UV continua. Combined with item (4) this is interpreted here as 
an inclination dependent reddening and absorption effect. These effects
must dominate radiation pressure effects in these radio sources, since the 
opposite correlations are otherwise expected.

(6) 
Absorbed radio-quiet quasars are also redder but stronger narrow absorption 
is detected in more luminous sources. This supports the picture in which radiation 
pressure is related to and perhaps responsible for the stronger absorption.
This would be expected if increased radiation pressure in more luminous
sources generate and accelerate more absorbing outflows by, for example, 
heating the upper layers of the accretion disk, as advocated in some disk
outflow models. 
Narrow absorption stronger than $EW_{\rm rest} \approx$ 3\AA{} is not
seen for the radio-quiet quasars.

(7)
In addition to a higher incidence of associated \civ{} absorbers a
clustering of absorbers is detected in both radio-loud and radio-quiet
quasars at outflow velocities around $-$15\,000\,\kms{} and an increase 
in absorber strength is seen around $-$18\,000\,\kms{} (Fig.~\ref{vel.fig}). 
This is intriguingly close to the common terminal velocity 
(about $-$20\,000\,\kms) seen for broad absorption line (BAL) outflows.  
It is here suggested that the intermediate strength absorbers 
($EW_{\rm rest} \lsim$ 3\AA) may constitute a slightly different 
population of narrow absorbers, perhaps entrained by BAL outflows in 
radio-quiet quasars and radio outflows in radio quasars.

(8) 
The dearth of UV narrow absorbers in ``face-on'' radio quasars 
(\ie core-dominated) reported by Ganguly \et (2001) for their sample 
of $z \lsim 1$ quasars is not supported. In contrast, a strikingly high 
fraction of $z \approx 2$ radio core-dominated quasars with emission line 
FWHM(\civ) $\lsim$6000\,\kms{} have detected absorption stronger than
$EW \geq$0.5\AA{} (Tables~\ref{fwfreq1.tab}; Fig.~\ref{fw_ewvel.fig}).
While redshift evolution 
of the column density and distribution of the absorbing gas may explain
the different results, the sample selection is suspected the main cause,
as the $z \lsim 0.5$ BQS quasars are as strongly absorbed as the quasars here.

(9) 
The strongest absorption is seen among the radio quasars with the largest
extent of the radio emission and the upper envelope in absorption
strength decreases with decreasing linear radio size 
(Fig.~\ref{linsize.fig}).  This contrasts
the results of Baker \et (2002). For the well defined, homogeneous, 
low-frequency selected (source inclination unbiased) Molonglo quasar 
sample they find the strongest narrow associated \civ{} \lam 1549 
absorption among the most dust-reddened quasars with the smallest 
linear extent of their radio emission, namely among the compact steep 
spectrum sources, believed to be young radio sources. Baker 
\et propose that the presence and strength of
narrow associated absorption is likely related to radio source growth
such that smaller (younger) radio sources are more strongly absorbed.
Although less homogeneous, the radio quasar sample studied here is well
selected with respect to source inclination and intrinsic brightness,
and so should not be strongly biased in this regard.  The deviating 
results are at present attributed to the fact that the radio quasars 
here are intrinsically more luminous by a factor $\geq$10 than 
the Molonglo quasars, possibly enabling stronger absorbing outflows.
However, the small number statistics of both studies may also contribute.

\bigskip
Items (4), (5), and (6) are consistent with expectations of disk-wind 
scenarios (\eg Murray \& Chiang 1995, 1997) in which the strong X-ray
flux in radio-loud quasars overionize the somewhat equatorially outflowing
wind to a high degree, thereby significantly reducing the effects
of the radiation pressure on the wind. In this case, the stronger
associated narrow absorption lines in radio-loud quasar are explained
as ``failed'' broad absorption line outflows. The lack of individual 
narrow associated absorption systems with $EW_{\rm rest} >$ 3\AA{}
and the higher frequency of BALs among radio-quiet quasars are also 
consistent with the purported stronger ability of RQQs to accelerate 
the outflows.

\acknowledgments

Many thanks are due to my collaborator and former co-advisor Peter Barthel 
for allowing me to include the radio-loud quasar data in this follow-up 
study to my thesis work.  
I thank Luis Ho for helpful discussions early in the course of this work,
and Peter Barthel for comments on the manuscript.
The anonymous referee is thanked for careful reading of the manuscript
and for helpful comments.
Financial support from the Columbus Fellowship is gratefully acknowledged.
Part of the observations reported here were obtained at the MMT Observatory, 
a joint facility of the Smithsonian Institution and the University of Arizona
This research has made use of the NASA/IPAC Extragalactic Database (NED)
which is operated by the Jet Propulsion Laboratory, California Institute
of Technology, under contract with the National Aeronautics and Space
Administration.


\clearpage

\clearpage

\setcounter{table}{0}

\begin{deluxetable}{cccccrccrrcc}
\tablewidth{590pt}
\rotate
\tablecaption{Properties of Absorbed Quasars and their Absorbers \label{dataobsabs.tab}}
\tablehead{
\colhead{Object} &
\colhead{Radio-} &
\colhead{$z$} &
\colhead{$M_V$} &
\colhead{$\log L_{\lambda}$(1550\AA)} &
\colhead{$\alpha_{\lambda}$} &
\colhead{$\lambda_{\rm obs}$ } &
\colhead{$z_{\rm abs}$} &
\colhead{$v$} &
\colhead{$\beta$} &
\colhead{$EW_{\rm abs}$} &
\colhead{$\sigma(EW)$}  \nl
\colhead{\nodata} &
\colhead{type} &
\colhead{\nodata} &
\colhead{[mag]} &
\colhead{[erg s$^{-1}$]} &
\colhead{\nodata} &
\colhead{[\AA ]} &
\colhead{\nodata} &
\colhead{[km s$^{-1}$]} &
\colhead{\nodata} &
\colhead{[\AA ]} &
\colhead{[\AA ]} \nl
\colhead{(1)} &
\colhead{(2)} &
\colhead{(3)} &
\colhead{(4)} &
\colhead{(5)} &
\colhead{(6)} &
\colhead{(7)} &
\colhead{(8)} &
\colhead{(9)} &
\colhead{(10)} &
\colhead{(11)} &
\colhead{(12)} 
}
\tablecolumns{12}
\startdata
Q0000$-$001  &  RQ  &  2.576  &  $-$27.18  &  43.430  &  $-1.30 \pm 0.14$ &  5543.8  &  2.5790  &  241  &  $-$0.0008  &  0.50 & 0.03\\ 
             &      &          &            &          &                   &  5512.5  &  2.5587  &  $-$1458  &  0.0049  &  0.91 & 0.03\\ 
Q0008$-$008  &  RQ  &  2.051  &  $-$27.18  &  43.201  &  $-1.42 \pm 0.21$ &  4600.6  &  1.9701  &  $-$8014  &  0.0267  &  0.94 & 0.05\\ 
Q0017$+$154  &  LDQ &  2.015  &  $-$28.30  &  43.994  &  $-1.64 \pm 0.09$ &  4450.7  &  1.8733  &  $-$14437  &  0.0482  &  1.26 & 0.09\\ 
             &      &          &            &          &                   &  4550.7  &  1.9378  &  $-$7789  &  0.0260  &  0.50 & 0.09\\ 
Q0106$+$013  &  CDQ &  2.100  &  $-$28.21  &  43.433  &  $-2.06 \pm 0.21$ &  4799.2  &  2.0983  &  $-$169  &  0.0006  &  0.46 & 0.02\\ 
Q0107$-$005  &  RQ  &  1.749  &  $-$27.25  &  43.079  &  $-1.92 \pm 0.48$ &  4264.3  &  1.7530  &  417  &  $-$0.0014  &  1.26 & 0.05\\ 
Q0109$+$176  &  LDQ &  2.152  &  $-$27.90  &  43.650  &  $-1.36 \pm 0.31$ &  4889.0  &  2.1562  &  452  &  $-$0.0015  &  4.77 & 0.06\\ 
             &      &          &            &          &                   &  4620.5  &  1.9829  &  $-$16466  &  0.0549  &  0.52 & 0.11\\ 
Q0225$-$014  &  LDQ &  2.039  &  $-$27.94  &  43.481  &  $-1.77 \pm 0.25$ &  4642.7  &  1.9972  &  $-$4174  &  0.0139  &  1.39 & 0.08\\ 
Q0226$-$038  &  CDQ &  2.060  &  $-$29.60  &  43.798  &  $-1.79 \pm 0.15$ &  4713.8  &  2.0431  &  $-$1631  &  0.0054  &  0.38 & 0.03\\ 
Q0244$+$017  &  RQ  &  1.948  &  $-$27.13  &  43.159  &  $-1.15 \pm 0.25$ &  4588.4  &  1.9622  &  1397  &  $-$0.0047  &  2.99 & 0.04\\ 
Q0252$+$016  &  RQ  &  2.457  &  $-$28.51  &  44.104  &  $-1.27 \pm 0.11$ &  5108.9  &  2.2982  &  $-$14120  &  0.0471  &  0.35 & 0.04\\ 
Q0253$-$024  &  RQ  &  1.986  &  $-$27.14  &  43.176  &  $-1.51 \pm 0.33$ &  4533.0  &  1.9264  &  $-$6001  &  0.0200  &  0.81 & 0.07\\ 
Q0254$-$016  &  RQ  &  2.685  &  $-$27.86  &  43.601  &  $-1.52 \pm 0.28$ &  5449.6  &  2.5181  &  $-$13852  &  0.0462  &  0.70 & 0.07\\ 
Q0258$+$021  &  RQ  &  2.521  &  $-$28.85  &  43.955  &  $-2.11 \pm 0.12$ &  5477.0  &  2.5358  &  1237  &  $-$0.0041  &  0.17 & 0.02\\ 
Q0317$-$023  &  CDQ &  2.076  &  $-$27.08  &  42.764  &  $-1.03 \pm 0.43$ &  4743.8  &  2.0625  &  $-$1289  &  0.0043  &  1.39 & 0.05\\ 
Q0348$+$061  &  RQ  &  2.058  &  $-$28.95  &  43.771  &  $-0.79 \pm 0.10$ &  4399.3  &  1.8401  &  $-$22093  &  0.0737  &  0.56 & 0.05\\ 
             &      &          &            &          &                   &  4595.5  &  1.9668  &  $-$9052  &  0.0302  &  0.69 & 0.05\\ 
             &      &          &            &          &                   &  4682.2  &  2.0227  &  $-$3450  &  0.0115  &  1.26 & 0.03\\ 
             &      &          &            &          &                   &  4696.0  &  2.0316  &  $-$2568  &  0.0086  &  0.76 & 0.03\\ 
Q0445$+$097  &  CSS &  2.106  &  $-$26.99  &  42.923  &  $-1.12 \pm 0.60$ &  4811.8  &  2.1064  &  16  &  $-$0.0001  &  4.43 & 0.08\\ 
             &      &          &            &          &                   &  4820.1  &  2.1118  &  533  &  $-$0.0018  &  2.22 & 0.08\\ 
Q0458$-$020  &  CDQ &  2.286  &  $-$27.03  &  43.431  &  $-0.20 \pm 0.18$ &  4706.0  &  2.0381  &  $-$23437  &  0.0782  &  1.50 & 0.11\\ 
             &      &          &            &          &                   &  4783.2  &  2.0879  &  $-$18585  &  0.0620  &  0.63 & 0.11\\ 
Q0504$+$030  &  LDQ &  2.470  &  $-$27.74  &  43.389  &  $-1.63 \pm 0.33$ &  5280.1  &  2.4087  &  $-$5364  &  0.0179  &  1.15 & 0.10\\ 
Q0751$+$298  &  CSS &  2.106  &  $-$28.10  &  44.054  &  $-2.33 \pm 0.21$ &  4519.7  &  1.9178  &  $-$18688  &  0.0623  &  0.69 & 0.08\\ 
Q0802$+$103  &  LDQ &  1.952  &  $-$28.11  &  43.939  &  $-1.80 \pm 0.20$ &  4568.4  &  1.9492  &  $-$289  &  0.0010  &  6.17 & 0.02\\ 
Q0805$+$046  &  CDQ &  2.876  &  $-$29.24  &  44.409  &  $-2.26 \pm 0.09$ &  6004.0  &  2.8761  &  32  &  $-$0.0001  &  0.27 & 0.02\\ 
Q0808$+$289  &  LDQ &  1.886  &  $-$27.99  &  43.858  &  $-1.17 \pm 0.14$ &  4307.1  &  1.7805  &  $-$11171  &  0.0373  &  0.45 & 0.06\\ 
             &      &          &            &          &                   &  4450.0  &  1.8728  &  $-$1391  &  0.0046  &  2.32 & 0.05\\ 
Q0831$+$101  &  LDQ &  1.757  &  $-$26.47  &  42.835  &  $-1.48 \pm 0.53$ &  4274.7  &  1.7596  &  284  &  $-$0.0009  &  0.88 & 0.09\\ 
             &      &          &            &          &                   &  4243.0  &  1.7392  &  $-$1947  &  0.0065  &  0.96 & 0.10\\ 
Q0835$+$580  &  LDQ &  1.528  &  $-$28.12  &  43.259  &  $-0.27 \pm 0.11$ &  3922.3  &  1.5321  &  530  &  $-$0.0018  &  1.24 & 0.07\\ 
             &      &          &            &          &                   &  3937.5  &  1.5420  &  1696  &  $-$0.0057  &  1.57 & 0.07\\ 
Q0941$+$261  &  CSS &  2.906  &  $-$29.58  &  43.840  &  $-2.67 \pm 0.29$ &  6064.2  &  2.9149  &  684  &  $-$0.0023  &  0.33 & 0.06\\ 
Q1016$-$006  &  RQ  &  2.181  &  $-$27.87  &  43.541  &  $-1.80 \pm 0.14$ &  4573.2  &  1.9524  &  $-$22355  &  0.0746  &  1.38 & 0.04\\ 
Q1023$+$067  &  LDQ &  1.698  &  $-$27.48  &  43.188  &  $-0.90 \pm 0.19$ &  4204.1  &  1.7141  &  1742  &  $-$0.0058  &  7.59 & 0.06\\ 
Q1045$+$052  &  RQ  &  2.120  &  $-$27.38  &  43.297  &  $-1.44 \pm 0.22$ &  4535.9  &  1.9283  &  $-$18946  &  0.0632  &  1.62 & 0.07\\ 
Q1116$+$128  &  CDQ &  2.124  &  $-$27.29  &  43.590  &  $-1.69 \pm 0.15$ &  4845.0  &  2.1279  &  399  &  $-$0.0013  &  1.91 & 0.03\\ 
             &      &         &            &          &                   &  4611.1  &  1.9768  &  $-$14424  &  0.0481  &  0.84 & 0.05\\ 
Q1158$+$122  &  LDQ &  2.014  &  $-$28.90  &  43.864  &  $-1.27 \pm 0.10$ &  4642.0  &  1.9968  &  $-$1683  &  0.0056  &  1.10 & 0.04\\ 
Q1208$+$105  &  RQ  &  1.864  &  $-$27.80  &  42.841  &  $-1.20 \pm 0.59$ &  4431.5  &  1.8609  &  $-$278  &  0.0009  &  1.66 & 0.10\\ 
Q1219$+$491  &  RQ  &  2.323  &  $-$27.70  &  43.484  &  $-1.62 \pm 0.18$ &  4929.3  &  2.1822  &  $-$12955  &  0.0432  &  0.58 & 0.08\\ 
             &      &         &            &          &                   &  5031.1  &  2.2480  &  $-$6834  &  0.0228  &  1.30 & 0.08\\ 
Q1223$+$178  &  RQ  &  2.930  &  $-$29.48  &  44.224  &  $-1.43 \pm 0.23$ &  5661.0  &  2.6546  &  $-$21715  &  0.0724  &  0.75 & 0.05\\ 
             &      &          &            &          &                   &  6037.4  &  2.8976  &  $-$2454  &  0.0082  &  1.12 & 0.04\\ 
             &      &          &            &          &                   &  6130.2  &  2.9575  &  2118  &  $-$0.0071  &  1.62 & 0.04\\ 
Q1226$+$105  &  LDQ &  2.303  &  $-$28.04  &  44.442  &  $-2.65 \pm 0.11$ &  4790.2  &  2.0924  &  $-$19695  &  0.0657  &  0.41 & 0.07\\ 
Q1227$+$120  &  RQ  &  2.440  &  $-$27.54  &  43.310  &  $-0.93 \pm 0.22$ &  5380.2  &  2.4734  &  2883  &  $-$0.0096  &  0.75 & 0.10\\ 
Q1230$+$164  &  RQ  &  2.716  &  $-$29.41  &  44.329  &  $-1.48 \pm 0.10$ &  5713.2  &  2.6883  &  $-$2256  &  0.0075  &  0.82 & 0.04 \\ 
             &      &          &            &          &                   &  5738.7  &  2.7048  &  $-$923  &  0.0031  &  2.37 & 0.03\\ 
Q1232$-$004  &  RQ  &  1.587  &  $-$26.65  &  42.859  &  $-1.70 \pm 0.29$ &  3837.4  &  1.4774  &  $-$12953  &  0.0432  &  0.92 & 0.08\\ 
Q1246$-$022  &  RQ  &  2.109  &  $-$28.49  &  43.911  &  $-1.43 \pm 0.08$ &  4473.4  &  1.8880  &  $-$22058  &  0.0736  &  0.93 & 0.04\\ 
Q1259$+$344  &  RQ  &  2.754  &  $-$28.33  &  43.752  &  $-0.99 \pm 0.14$ &  5706.7  &  2.6841  &  $-$5670  &  0.0189  &  1.13 & 0.04\\ 
             &      &          &            &          &                   &  5780.6  &  2.7318  &  $-$1813  &  0.0060  &  2.15 & 0.03\\ 
             &      &          &            &          &                   &  5841.0  &  2.7708  &  1306  &  $-$0.0044  &  3.44 & 0.03\\ 
Q1313$+$200  &  CDQ &  2.462  &  $-$28.26  &  43.985  &  $-1.88 \pm 0.08$ &  5373.0  &  2.4687  &  570  &  $-$0.0019  &  1.98 & 0.03\\ 
Q1323$+$655  &  LDQ &  1.624  &  $-$28.15  &  44.538  &  $-2.88 \pm 0.16$ &  4043.3  &  1.6102  &  $-$1527  &  0.0051  &  0.41 & 0.05\\ 
Q1402$-$012  &  CDQ &  2.518  &  $-$28.67  &  43.968  &  $-2.23 \pm 0.16$ &  5226.9  &  2.3744  &  $-$12447  &  0.0415  &  0.37 & 0.08\\ 
Q1409$+$095  &  RQ  &  2.840  &  $-$28.89  &  43.933  &  $-2.19 \pm 0.38$ &  5679.2  &  2.6663  &  $-$13839  &  0.0462  &  0.72 & 0.07\\ 
             &      &          &            &          &                   &  5762.8  &  2.7203  &  $-$9465  &  0.0316  &  0.87 & 0.06\\ 
Q1434$-$009  &  RQ  &  1.669  &  $-$27.71  &  43.432  &  $-1.73 \pm 0.14$ &  3966.8  &  1.5609  &  $-$12408  &  0.0414  &  0.56 & 0.05\\ 
             &      &          &            &          &                   &  3969.1  &  1.5624  &  $-$12231  &  0.0408  &  0.56 & 0.05\\ 
Q1443$-$010  &  RQ  &  1.795  &  $-$27.82  &  43.594  &  $-0.95 \pm 0.13$ &  4308.3  &  1.7814  &  $-$1421  &  0.0047  &  1.67 & 0.04\\ 
Q1540$+$180  &  LDQ &  1.663  &  $-$27.90  &  43.277  &  $-2.46 \pm 0.33$ &  3940.1  &  1.5436  &  $-$13679  &  0.0456  &  0.68 & 0.10\\ 
Q1606$+$289  &  LDQ &  1.983  &  $-$27.46  &  43.326  &  $-1.15 \pm 0.21$ &  4600.0  &  1.9697  &  $-$1372  &  0.0046  &  9.11 & 0.16\\ 
Q1614$+$051  &  GPS &  3.212  &  $-$27.73  &  43.850  &  $-1.73 \pm 0.10$ &  6344.3  &  3.0957  &  $-$8400  &  0.0280  &  0.41 & 0.04\\ 
             &      &          &            &          &                   &  6128.9  &  2.9567  &  $-$18732  &  0.0625  &  0.41 & 0.04\\ 
Q1629$+$120  &  CSS &  1.781  &  $-$27.63  &  43.306  &  $-2.09 \pm 0.32$ &  4363.3  &  1.8169  &  3869  &  $-$0.0129  &  0.81 & 0.07\\ 
Q1629$+$680  &  CSS &  2.473  &  $-$27.78  &  43.581  &  $-2.44 \pm 0.20$ &  5401.6  &  2.4872  &  1255  &  $-$0.0042  &  0.33 & 0.04\\ 
Q1633$+$382  &  CDQ &  1.810  &  $-$28.06  &  44.056  &  $-0.87 \pm 0.05$ &  4096.3  &  1.6445  &  $-$18133  &  0.0605  &  0.30 & 0.02\\ 
Q1656$+$477  &  CDQ &  1.614  &  $-$27.84  &  43.804  &  $-1.86 \pm 0.18$ &  3895.4  &  1.5148  &  $-$11585  &  0.0386  &  0.77 & 0.05\\ 
Q1658$+$575  &  LDQ &  2.171  &  $-$28.38  &  43.640  &  $-1.93 \pm 0.13$ &  4931.8  &  2.1839  &  1196  &  $-$0.0040  &  0.27 & 0.03\\ 
Q1701$+$379  &  LDQ &  2.454  &  $-$27.75  &  43.372  &  $-2.52 \pm 0.40$ &  5311.6  &  2.4291  &  $-$2187  &  0.0073  &  0.52 & 0.07\\ 
             &      &          &            &          &                   &  4909.0  &  2.1692  &  $-$25753  &  0.0859  &  0.81 & 0.11\\ 
Q1726$+$344  &  CSS &  2.426  &  $-$28.20  &  43.768  &  $-2.05 \pm 0.08$ &  5204.6  &  2.3600  &  $-$6000  &  0.0200  &  2.00 & 0.05\\ 
Q2048$+$196  &  CDQ &  2.364  &  $-$28.11  &  43.620  &  $-1.84 \pm 0.15$ &  5193.3  &  2.3527  &  $-$1049  &  0.0035  &  1.07 & 0.04\\ 
Q2150$+$053  &  LDQ &  1.979  &  $-$28.55  &  43.767  &  $-1.33 \pm 0.12$ &  4630.7  &  1.9895  &  1013  &  $-$0.0034  &  0.62 & 0.03\\ 
Q2222$+$051  &  LDQ &  2.322  &  $-$28.06  &  43.754  &  $+0.51 \pm 0.28$ &  5129.6  &  2.3115  &  $-$973  &  0.0032  &  1.23 & 0.04\\ 
             &      &          &            &          &                   &  4848.3  &  2.1300  &  $-$17857  &  0.0596  &  2.38 & 0.06\\ 
Q2251$+$244  &  CSS &  2.313  &  $-$28.36  &  43.773  &  $-1.41 \pm 0.12$ &  4887.9  &  2.1555  &  $-$14601  &  0.0487  &  1.47 & 0.07\\ 
             &      &          &            &          &                   &  5192.3  &  2.3520  &  3498  &  $-$0.0117  &  0.92 & 0.06\\ 
             &      &          &            &          &                   &  5210.6  &  2.3639  &  4554  &  $-$0.0152  &  2.35 & 0.07\\ 
Q2351$+$022  &  RQ  &  2.024  &  $-$27.91  &  43.473  &  $-1.34 \pm 0.23$ &  4702.7  &  2.0360  &  1158  &  $-$0.0039  &  2.26 & 0.06\\ 
Q2359$+$002  &  RQ  &  2.675  &  $-$27.35  &  43.475  &  $-2.98 \pm 0.24$ &  5697.5  &  2.6782  &  291  &  $-$0.0010  &  2.42 & 0.04\\ 
\enddata
\tablecomments{
col (1): Quasar name;
col (2): Radio type: RQ= radio quiet quasar, CDQ= core-dominated quasar, LDQ= lobe-dominated quasar, 
	   CSS= compact steep-spectrum source, GPS = Giga-hertz Peaked spectrum source (see \S~2); 
col (3): Quasar redshift;
col (4): Absolute V-band magnitude ($\alpha = -0.7$ was adopted for the K-correction; $F_{\lambda} \propto \lambda^{\alpha_{\lambda}}$);
col (5): logarithm of the monochromatic continuum luminosity at 1550\AA;
col (6): UV spectral slope, $\alpha_{\lambda}$;
col (7): Wavelength position of the narrow absorber in the observed frame of reference;
col (8): Redshift of the absorber;
col (9): Absorber velocity defined here as  $v = -\beta c$. That is, negative velocities
denote blueshifted absorbers relative to the emission line redshift. Since UV broad emission lines can
be blueshifted relative to the systemic redshift, redshifted absorption need not indicate infall toward
the central engine;
col (10): $\beta =(r^2 - 1)/(r^2 + 1)$ where $r = (1 + z_{\rm em})/(1 + z_{\rm abs})$, and 
          $c$ is the speed of light (Weymann et al.\ 1979; Peterson 1997);
col (11): Restframe equivalent width of the narrow absorber;
col (12): The 1\,$\sigma$ $EW$ uncertainty measured from the continuum $rms$ across a resolution element 
       at the position of the absorber.
}
\end{deluxetable}

%

\begin{deluxetable}{cccccr}  
\tablewidth{305pt}  
\tabletypesize{\small} 
\tablecaption{Properties of Unabsorbed Quasars \label{dataObsNoAbs.tab}}
\tablehead{  
\colhead{Object} &  
\colhead{Radio-} &  
\colhead{$z$} & 
\colhead{$M_V$} &
\colhead{$\log L_{\lambda}$(1550\AA)} &
\colhead{$\alpha_{\lambda}$} \nl
\colhead{\nodata} & 
\colhead{type} &
\colhead{\nodata} & 
\colhead{[mag]} &
\colhead{[erg s$^{-1}$]} &
\colhead{\nodata} \nl
\colhead{(1)} &
\colhead{(2)} &
\colhead{(3)} &
\colhead{(4)} &
\colhead{(5)} &
\colhead{(6)} 
}
\tablecolumns{6}
\startdata
Q0002$-$008  &  RQ  &  2.166  &  $-$28.56  &  43.561  &  $-2.47 \pm 0.26$  \\ 
Q0003$-$006  &  RQ  &  1.723  &  $-$26.46  &  43.127  &  $-1.67 \pm 0.20$  \\ 
Q0015$+$026  &  RQ  &  2.468  &  $-$28.06  &  43.428  &  $-1.62^{~+~0.42~}_{~-~0.50~}$ \\
Q0020$+$022  &  RQ  &  1.788  &  $-$27.53  &  43.362  &  $-2.27 \pm 0.20$  \\ 
Q0038$-$019  & LDQ  &  1.672  &  $-$27.42  &  43.893  &  $-2.22 \pm 0.14$  \\ 
Q0040$-$017  &  RQ  &  2.398  &  $-$28.70  &  44.159  &  $-1.90 \pm 0.25$  \\ 
Q0115$-$011  &  RQ  &  2.179  &  $-$28.28  &  43.518  &  $-1.89 \pm 0.35$  \\ 
Q0206$+$293  & CDQ  &  2.197  &  $-$27.54  &  43.169  &  $-1.88 \pm 0.23$  \\ 
Q0238$+$100  & LDQ  &  1.829  &  $-$28.19  &  43.587  &  $-1.54^{~+~0.40~}_{~-~0.51~}$ \\
Q0352$+$123  & LDQ  &  1.616  &  $-$26.53  &  43.166  &  $-1.11 \pm 0.29$  \\ 
Q0730$+$257  & LDQ  &  2.688  &  $-$27.20  &  43.787  &  $-2.12 \pm 0.30$ \\ 
Q0758$+$120  & LDQ  &  2.658  &  $-$27.16  &  43.677  &  $-2.56 \pm 0.22$  \\ 
Q0856$+$124  & CSS  &  1.767  &  $-$26.69  &  43.058  &  $-2.35^{~+~0.47~}_{~-~0.63~}$ \\
Q0926$+$117  & LDQ  &  1.754  &  $-$26.96  &  43.698  &  $-2.17 \pm 0.16$ \\ 
Q1020$+$014  &  RQ  &  1.605  &  $-$27.70  &  43.386  &  $-1.57 \pm 0.25$  \\ 
Q1043$+$071  &  RQ  &  2.115  &  $-$27.3   &  43.187  &  $-1.68 \pm 0.25$  \\ 
Q1046$+$058  &  RQ  &  1.971  &  $-$27.74  &  43.470  &  $-1.67 \pm 0.15$  \\ 
Q1055$+$499  & CSS  &  2.387  &  $-$26.65  &  43.551  &  $-2.29 \pm 0.25$  \\ 
Q1137$+$305  &  RQ  &  1.584  &  $-$28.89  &  43.971  &  $-1.60 \pm 0.06$  \\ 
Q1138$+$002  &  RQ  &  1.760  &  $-$27.46  &  43.147  &  $-1.81 \pm 0.19$ \\ 
Q1146$+$111  &  RQ  &  1.932  &  $-$27.46  &  43.448  &  $-1.40 \pm 0.09$  \\ 
Q1214$+$106  & LDQ  &  1.881  &  $-$27.79  &  43.450  &  $-2.64^{~+~0.58~}_{~-~0.84~}$ \\
Q1221$+$113  & CSS  &  1.759  &  $-$27.28  &  44.008  &  $-2.37 \pm 0.14$  \\ 
Q1237$+$134  &  RQ  &  1.722  &  $-$28.21  &  43.727  &  $-1.33 \pm 0.08$  \\ 
Q1258$+$404  & LDQ  &  1.667  &  $-$26.51  &  43.142  &  $-2.41^{~+~0.48~}_{~-~0.64~}$ \\
Q1311$-$270  & LDQ  &  2.196  &  $-$29.14  &  44.305  &  $-2.17 \pm 0.19$  \\ 
Q1330$+$011  &  RQ  &  3.501  &  $-$29.34  &  43.863  & $-3.86^{~+~0.39~}_{~-~0.51~}$ \\
Q1402$+$044  & CDQ  &  3.205  &  $-$28.43  &  43.678  &  $-2.20 \pm 0.24$  \\ 
Q1440$-$004  &  RQ  &  1.811  &  $-$27.45  &  43.568  &  $-1.61 \pm 0.14$  \\ 
Q1442$+$101  & GPS  &  3.527  &  $-$30.30  &  44.376  &  $-1.24 \pm 0.09$  \\ 
Q1517$+$239  &  RQ  &  1.833  &  $-$27.78  &  43.497  &  $-1.26 \pm 0.14$  \\ 
Q1542$+$042  & CDQ  &  2.182  &  $-$28.56  &  43.701  &  $-1.59 \pm 0.32$  \\ 
Q1554$-$203  & LDQ  &  1.938  &  $-$27.20  &  43.464  &  $-1.88 \pm 0.27$  \\ 
Q1556$-$245  & CDQ  &  2.810  &  $-$28.23  &  43.631  &  $-1.52 \pm 0.26$  \\ 
Q1602$+$576  & CSS  &  2.854  &  $-$29.20  &  44.781  &  $-3.31 \pm 0.15$  \\ 
Q1607$+$183  & CDQ  &  3.097  &  $-$28.84  &  43.941  &  $-2.33 \pm 0.23$  \\ 
Q1634$+$406  &  RQ  &  1.730  &  $-$27.38  &  43.313  &  $-2.36^{~+~0.36~}_{~-~0.46~}$ \\
Q1638$+$390  &  RQ  &  2.374  &  $-$28.11  &  43.736  &  $-1.43 \pm 0.14$  \\ 
Q1702$+$298  & CSS  &  1.928  &  $-$27.27  &  42.690  &  $-2.00 \pm 0.00$ \\                             
Q1704$+$710  &  RQ  &  2.010  &  $-$28.99  &  43.788  &  $-1.06 \pm 0.11$  \\ 
Q1705$+$018  & GPS  &  2.571  &  $-$28.06  &  43.540  &  $-2.21^{~+~0.50~}_{~-~0.67~}$ \\
Q1816$+$475  & LDQ  &  2.221  &  $-$28.30  &  43.792  &  $-2.41 \pm 0.20$  \\ 
Q1857$+$566  & LDQ  &  1.576  &  $-$28.48  &  43.663  &  $-2.33 \pm 0.12$  \\ 
Q2158$+$101  & CSS  &  1.727  &  $-$28.32  &  43.328  &  $-1.70 \pm 0.18$  \\ 
Q2212$-$299  & CDQ  &  2.702  &  $-$29.92  &  44.341  &  $-2.46 \pm 0.07$  \\ 
Q2223$+$210  & CDQ  &  1.947  &  $-$28.61  &  43.975  &  $-2.53 \pm 0.19$  \\ 
Q2233$+$136  &  RQ  &  3.210  &  $-$27.22  &  43.812  &  $-4.25 \pm 0.09$ \\
Q2239$+$007  &  RQ  &  2.412  &  $-$26.72  &  43.391  &  $-2.00 \pm 0.19$  \\ 
Q2248$+$192  & LDQ  &  1.794  &  $-$27.64  &  43.236  &  $-1.60 \pm 0.28$  \\ 
Q2334$+$019  &  RQ  &  2.185  &  $-$27.54  &  43.705  &  $-1.36 \pm 0.18$  \\ 
Q2341$+$010  &  RQ  &  2.305  &  $-$27.00  &  42.888  &  $-1.65^{~+~0.64~}_{~-~0.77~}$ \\
Q2345$+$061  & CSS  &  1.538  &  $-$27.93  &  43.384  &  $-1.76 \pm 0.16$  \\ 
Q2350$-$007  &  RQ  &  1.617  &  $-$27.23  &  43.228  &  $-0.70 \pm 0.20$  \\ 
\enddata
\tablecomments{colums (1) $-$ (6): see notes to Table~1.}
\end{deluxetable}


\begin{table}[ht]
\small
\vspace{-1cm}
\begin{center}
\caption{Frequency of C\,{\small IV} NAL Absorbed Objects$^a$
\label{freq.tab}}
\tablewidth{360pt}
\vspace{-0.3cm}
\begin{tabular}{lccccc}
\\
 \tableline \tableline\\[-8pt]
&No.\ &\multicolumn{2}{c}{No. of Absorbed}& No. of Quasars & No. of Quasars \\
&
\multicolumn{1}{c}{of}&\multicolumn{2}{c}{Quasars with } &{with $|$Absorber }&{with 5000 $<$ } \\
{       }&
\multicolumn{1}{c}{QSOs}&\multicolumn{2}{c}{$|$Abs.~Vel.$|$ $\leq$} &{Velocity$| \leq$}&{$|$Absorber Velocity$|$} \\
{Sample}&
\multicolumn{1}{c}{}&\multicolumn{2}{c}{~21000 km s$^{-1}$}&{~~5000 km s$^{-1}$}&{$\leq$ ~21000 km s$^{-1}$} \\
\multicolumn{1}{c}{(1)} &\multicolumn{1}{c}{(2)} & \multicolumn{2}{c}{(3)} & (4) & (5)\\
\tableline \\[-8pt]
& & \multicolumn{1}{c}{\underline{~~~~~~All $EW$s~~~~~}} & \multicolumn{3}{c}{\underline{~~~~~~~~~~~~~~~~~~~~~~Doublet $EW$ $\geq$ 0.5\AA{}$^b$~~~~~~~~~~~~~~~~~~~~~~~~~~~}} \\
All QSOs     & 114 & 59$\pm$7.7; 52$\pm$\,~7\% & 45$\pm$6.7; 39$\pm$\,~5\% & 31$\pm$5.6; 27$\pm$\,~5\% & 21$\pm$4.6; 18$\pm$\,~4\% \\
RQQ          & ~48 & 22$\pm$4.7; 46$\pm$10\%   & 19$\pm$4.4; 40$\pm$\,~9\% & 12$\pm$3.5; 25$\pm$\,~7\% & 10$\pm$3.2; 21$\pm$\,~7\% \\
RLQ          & ~66 & 37$\pm$6.1; 56$\pm$\,~9\% & 26$\pm$5.1; 39$\pm$\,~7\% & 19$\pm$4.4; 29$\pm$\,~7\% & 11$\pm$3.3; 17$\pm$\,~5\% \\
CDQ          & ~18 & 11$\pm$3.3; 61$\pm$18\%   & ~6$\pm$2.4; 33$\pm$13\%   & ~4$\pm$2.0; 22$\pm$11\%   & ~3$\pm$1.7; 17$\pm$10\% \\
LDQ          & ~31 & 18$\pm$4.2; 58$\pm$14\%   & 15$\pm$3.9; 48$\pm$12\%   & 12$\pm$3.5; 80$\pm$23\%   & ~5$\pm$2.2; 33$\pm$15\% \\
CSS          & ~14 & ~7$\pm$2.6; 50$\pm$19\%   & ~5$\pm$2.2; 36$\pm$16\%   & ~3$\pm$1.7; 21$\pm$12\%   & ~3$\pm$1.7; 21$\pm$12\% \\
CSS$+$GPS    & ~17 & ~8$\pm$2.8; 47$\pm$17\%   & ~5$\pm$2.2; 29$\pm$13\%   & ~3$\pm$1.7; 18$\pm$10\%   & ~3$\pm$1.7; 18$\pm$10\%\\[3pt] 
\tableline \\[-8pt]
& &  \multicolumn{4}{c}{\underline{~~~~~~~~~~~~~~~~~~~~~~~~~~Strongly Absorbed Quasars$^c$~~~~~~~~~~~~~~~~~~~~~~~~~~~~~~~~~~}} \\
All QSOs     & 114 & \multicolumn{2}{c}{40$\pm$6.3; 35$\pm$\,~5\%} & 28$\pm$5.3; 25$\pm$\,~5\% & 10$\pm$3.2; ~9$\pm$\,~3\% \\
RQQ          & ~48 & \multicolumn{2}{c}{15$\pm$3.9; 31$\pm$\,~8\%} & 11$\pm$3.3; 23$\pm$\,~7\% & ~5$\pm$2.2; 10$\pm$\,~5\% \\
RLQ          & ~66 & \multicolumn{2}{c}{25$\pm$5.0; 38$\pm$\,~8\%} & 17$\pm$4.1; 26$\pm$\,~6\% & ~5$\pm$2.2; ~8$\pm$\,~3\% \\
CDQ          & ~18 & \multicolumn{2}{c}{~5$\pm$2.2; 28$\pm$12\%} & ~4$\pm$2.0; 22$\pm$11\% & ~0$\pm$0.0; ~0$\pm$\,~0\% \\
LDQ          & ~31 & \multicolumn{2}{c}{14$\pm$3.7; 45$\pm$12\%} & 10$\pm$3.2; 32$\pm$10\% & ~3$\pm$1.7; 10$\pm^{\,~5}_{\,~6}$\% \\
CSS          & ~14 & \multicolumn{2}{c}{~5$\pm$2.2; 36$\pm$16\%} & ~2$\pm$1.4; 14$\pm$10\% & ~2$\pm$1.4; 14$\pm$10\% \\
CSS$+$GPS    & ~17 & \multicolumn{2}{c}{~5$\pm$2.2; 29$\pm$13\%} & ~2$\pm$1.4; 12$\pm$\,~8\% & ~2$\pm$1.4; 12$\pm$\,~8\% \\[3pt] 
\tableline \\[-8pt]
& &  \multicolumn{4}{c}{\underline{~~~~~~~~~~~~~~~~~~~~~~~~~~~Weakly Absorbed Quasars$^d$ ~~~~~~~~~~~~~~~~~~~~~~~~~~~~~~~~}} \\
All QSOs     & 114 & \multicolumn{2}{c}{24$\pm$4.9; 21$\pm$\,~4\%} & 11$\pm$3.3; 10$\pm$\,~3\% & 13$\pm$3.6; 11$\pm$\,~4\% \\
RQQ          & ~48 & \multicolumn{2}{c}{~7$\pm$2.6; 15$\pm$\,~5\%} & ~2$\pm$1.4; ~4$\pm$\,~3\% & ~5$\pm$2.2; 10$\pm$\,~5\% \\
RLQ          & ~66 & \multicolumn{2}{c}{17$\pm$4.1; 26$\pm$\,~6\%} & ~9$\pm$3.0; 14$\pm$\,~4\% & ~8$\pm$2.8; 12$\pm$\,~4\% \\
CDQ          & ~18 & \multicolumn{2}{c}{~7$\pm$2.6; 39$\pm$14\%} & ~3$\pm$1.7; 17$\pm$10\% & ~4$\pm$2.0; 22$\pm$11\% \\
LDQ          & ~31 & \multicolumn{2}{c}{~5$\pm$2.2; 16$\pm$\,~7\%} & ~3$\pm$1.7; 10$\pm$\,~6\% & ~2$\pm$1.4; ~6$\pm^{\,~5}_{\,~4}$\% \\
CSS          & ~14 & \multicolumn{2}{c}{~4$\pm$2.0; 29$\pm$14\%} & ~3$\pm$1.7; 21$\pm$13\% & ~1$\pm$1.0; ~7$\pm$\,~7\% \\
CSS$+$GPS    & ~17 & \multicolumn{2}{c}{~5$\pm$2.2; 29$\pm$13\%} & ~3$\pm$1.7; 18$\pm$10\% & ~2$\pm$1.4; 12$\pm$\,~8\% \\[3pt] 
\tableline
\tableline
\end{tabular}
\end{center}
\begin{center}
\begin{minipage}{15cm}
\footnotetext{ }{$a$ Number of absorbed quasars and counting errors. The frequency of absorbed quasars \\
include the uncertainties corresponding to the counting errors.} \\
\footnotetext{ }{$b$ The most complete subset of absorbed quasars have individual \civ{} doublet absorbers
with $EW_{\rm rest} \geq$0.5\AA{} and absolute velocity $\leq$ 21\,000\,\kms. See text and Figure~\ref{ewdistrib.fig}.}\\
\footnotetext{ }{$c$ For each velocity bin, the summed absorption $EW_{\rm rest}$ for each quasar must be at
least 1\AA{} to be counted. An individual quasar can have strong high and low velocity 
absorption and so be counted twice; this is, however, only seen in a couple of cases.} \\
\footnotetext{ }{$d$ The summed absorption $EW_{\rm rest}$ for each quasar must be less than 1\AA{} 
to be counted. This is mainly for comparison with studies finding only weak absorbers. Note that individual
absorbers with $EW <$\,0.5\AA{} are included here.}
\end{minipage}
\end{center}
\end{table}

\vspace{0.50cm}
\begin{table}[h]
\begin{center}
\caption{Absorber Frequency Among Absorbed Quasars Only
\label{freq3.tab}}
\begin{tabular}{lcccccc}
\\
 \tableline \tableline\\[-8pt]
{ }&\multicolumn{2}{c}{Number of Ab-} &{\#Absrs$^a$ for }&{\#Absrs$^a$ for }&{\#Absrs$^a$ for 5000 $<$ } \\
{ }&\multicolumn{2}{c}{sorbers$^a$ (\#Absrs)} &{$|$Abs.~Velocity$|$} &{$|$Abs.~Velocity$|$} &{ $|$Absorber Velocity$|$} \\
{Sample}&\multicolumn{2}{c}{(All Abs.\,Vel's)}&{$\leq$~21000 km s$^{-1}$}&{$\leq$~5000 km s$^{-1}$}&{$\leq$~21000 km s$^{-1}$} \\
\multicolumn{1}{c}{(1)}&\multicolumn{2}{c}{(2)} & (3) & (4) & (5) \\
\tableline \\[-8pt]
& \multicolumn{1}{c}{\underline{~All $EW$s~}} & \multicolumn{4}{c}{\underline{~~~~~~~~~~~~~~~~~~~~~~~~~~~~~~Doublet $EW$ $\geq$ 0.5\AA{}$^b$~~~~~~~~~~~~~~~~~~~~~~~~~~~~~}} \\
All QSOs     & 86$\pm$9.3 & 71$\pm$8.4 & 65$\pm$8.1 & 40$\pm$6.3; ~62$\pm$\,~9\% & 25$\pm$5.0; 38$\pm$\,~7\% \\
RQQ          & 36$\pm$6.0 & 34$\pm$5.8 & 30$\pm$5.5 & 17$\pm$4.1; ~57$\pm$13\% & 13$\pm$3.6; 43$\pm$12\% \\
RLQ          & 50$\pm$7.1 & 37$\pm$6.1 & 35$\pm$5.9 & 23$\pm$4.8; ~66$\pm$13\% & 12$\pm$3.5; 34$\pm$10\% \\
CDQ          & 13$\pm$3.6 & ~8$\pm$2.8 & ~7$\pm$2.6 & ~4$\pm$2.0; ~57$\pm$28\% & ~3$\pm$1.7; 43$\pm$24\% \\
LDQ          & 25$\pm$5.0 & 21$\pm$4.6 & 20$\pm$4.5 & 14$\pm$3.7; ~70$\pm$18\% & ~6$\pm$2.4; 30$\pm$12\% \\
CSS          & 10$\pm$3.2 & ~8$\pm$2.8 & ~8$\pm$2.8 & ~5$\pm$2.2; ~63$\pm$28\% & ~3$\pm$1.7; 38$\pm$21\% \\
CSS$+$GPS    & 12$\pm$3.5 & ~8$\pm$2.8 & ~8$\pm$2.8 & ~5$\pm$2.2; ~63$\pm$28\% & ~3$\pm$1.7; 38$\pm$21\% \\[3pt] 
\tableline
& \multicolumn{5}{c}{\underline{~~~~~~~~~~~~~~~~~~~~~~~~~~~~~~~~~~~~~~~Strong Absorbers$^c$~~~~~~~~~~~~~~~~~~~~~~~~~~~~~~~~~~~~~~~}} \\
All QSOs     & \multicolumn{2}{c}{39$\pm$6.2}& 37$\pm$6.1 & 29$\pm$5.4; ~78$\pm$14\% & ~8$\pm$2.8; 22$\pm$\,~8\% \\
RQQ          & \multicolumn{2}{c}{16$\pm$4.0}& 15$\pm$3.9 & 12$\pm$3.5; ~80$\pm$23\% & ~3$\pm$1.7; 20$\pm$11\% \\
RLQ          & \multicolumn{2}{c}{23$\pm$4.8}& 22$\pm$4.7 & 17$\pm$4.1; ~77$\pm$18\% & ~5$\pm$2.2; 23$\pm$10\% \\
CDQ          & \multicolumn{2}{c}{~5$\pm$2.2}& ~4$\pm$2.0 & ~4$\pm$2.0; 100$\pm$50\% & ~0$\pm$0.0; ~0$\pm$\,~0\% \\
LDQ          & \multicolumn{2}{c}{13$\pm$3.6}& 13$\pm$3.6 & 10$\pm$3.2; ~77$\pm^{32}_{18}$\% & ~3$\pm$1.7; ~23$\pm$17\% \\
CSS          & \multicolumn{2}{c}{~5$\pm$2.2}& ~5$\pm$2.2 & ~3$\pm$1.7; ~60$\pm$34\% & ~2$\pm$1.4; 40$\pm$28\% \\
CSS$+$GPS    & \multicolumn{2}{c}{~5$\pm$2.2}& ~5$\pm$2.2 & ~3$\pm$1.7; ~60$\pm$34\% & ~2$\pm$1.4; 40$\pm$28\% \\[3pt] 
\tableline
\tableline
\end{tabular}
\end{center}
\begin{center}
\begin{minipage}{15cm}
\footnotetext{ }{$a$ Number of absorbers and counting errors. The frequency of absorbers include
the uncertainties corresponding to the counting errors.} \\
\footnotetext{ }{$b$ Absorbers with $EW_{\rm rest} \geq$0.5\AA{} reach the highest completeness level 
($\sim$95\%); see \S~\ref{compltness} and Figure~\ref{ewdistrib.fig}.}\\
\footnotetext{ }{$c$ Strong absorbers only: summed restframe $EW$ per quasar $\geq$ 1.0\AA.} 
\end{minipage}
\end{center}
\end{table}

\begin{table}[h]
\begin{center}
\caption{Correlation Test Statistics 
\label{stats.tab}}
\begin{tabular}{lclcccc}
\\
 \tableline\\[-8pt]
&&&\multicolumn{2}{c}{\underline{~~~~~~~~All NALs~~~~~~}}&\multicolumn{2}{c}{\underline{~~~~~~NAL $EW \geq 0.5$\AA{}$^c$~~~}}\\
{Independent}& {Dependent} & & \multicolumn{1}{c}{Spearman's} & \multicolumn{1}{c}{$P^b$}
& \multicolumn{1}{c}{Spearman's}&\multicolumn{1}{c}{$P$}\\
{Parameter}&{Parameter}& { Sample$^a$}& \multicolumn{1}{c}{$r$} & \multicolumn{1}{c}{(\%)}
&\multicolumn{1}{c}{$r$}&\multicolumn{1}{c}{(\%)}\\
\multicolumn{1}{c}{(1)}  & (2) &\multicolumn{1}{c}{(3)} &\multicolumn{1}{c}{(4)}
&\multicolumn{1}{c}{(5)}&\multicolumn{1}{c}{(6)}&\multicolumn{1}{c}{(7)}\\
\tableline \\[-8pt]
$EW^d$ & $\alpha_{\rm UV,\lambda}$ & All QSOs &  ~~0.53 &$<0.01$& ~~0.45 & ~0.2\\
       &                           & RQQs     &  ~~0.32 & 12.9 &  ~~0.33 & 13.0\\
       &                           & RLQs     &  ~~0.62 & ~0.02&  ~~0.56 & ~0.5 \\
$EW$   & $M_{V}$                   & All QSOs &  ~~0.23 & ~8.2 & $-$0.12 & 43.2\\
       &                           & RQQs     & $-$0.23 & 27.5 & $-$0.43 & ~4.9\\
       &                           & RLQs     &  ~~0.47 & ~0.5 &  ~~0.21 & 29.5\\
$EW$   & $L_{\rm cont}$            & All QSOs & $-$0.24 & ~6.4 &  ~~0.13 & 38.9\\
       &                           & RQQs     &  ~~0.15 & 48.4 &  ~~0.40 & ~7.1\\
       &                           & RLQs     & $-$0.41 & ~1.5 & $-$0.15 & 46.1\\
\tableline
\tableline
\end{tabular}
\end{center}
\begin{center}
\begin{minipage}{14cm}
$a$ There are 24 absorbed RQQs (22 with $EW \geq 0.5$\AA{}) and 37 absorbed RLQs (26 with $EW \geq 0.5$\AA{}).\\
$b$ Probability that no correlation is present. \\
$c$ Here only quasars with $EW \geq 0.5$\AA{} NALs are analyzed. \\
$d$ This is the total $EW$ for each quasar.\\
\end{minipage}
\end{center}
\end{table}

%
\begin{table}[h]
\begin{center}
\caption{Sample Comparisons$^a$ \label{qsamples.tab}}
\begin{tabular}{lcccc}
\\
 \tableline\\[-8pt]
\multicolumn{1}{c}{Sample} & $M_V$ range & $<M_V>$ & $\log$ [$L_{\nu}$(408MHz) & $\log$ [$< L_{\nu}$(408MHz)$>$ \\
& (mag) & (mag) & /W Hz$^{-1}$] range & /W Hz$^{-1}$] \\
\multicolumn{1}{c}{(1)}  & (2) &\multicolumn{1}{c}{(3)} &\multicolumn{1}{c}{(4)}
&\multicolumn{1}{c}{(5)}\\
\tableline \\[-8pt]
Current RLQ sample     & $-$25.5 to $-28.2$ & $-26.7$ & 28.9 to 30.3 & 29.6 \\
Baker$^b$ \et (2002)   & $-$23.0 to $-28.2$ & $-25.3$ & 28.1 to 29.0 & 28.5\\
Ganguly \et (2001)     & $-$23.5 to $-28.1$ & $-26.2$ & .... & ....\\
\tableline
\tableline
\end{tabular}
\end{center}
\begin{center}
\begin{minipage}{15cm}
$a$ Cosmology adopted: $H_0$ = 50 ${\rm km~ s^{-1} Mpc^{-1}}$, q$_0$ = 0.5, and $\Lambda$ = 0.
\\
$b$ The data presented by Kapahi \et (1998) were used
to compute the luminosities for the high-$z$ quasar sample studied by Baker
\et (2002).  The $B_J$ magnitudes were converted to $V$ magnitudes as follows: 
the color correction $B - B_J = 0.24 (B-V)$ of Evans (1989) combined with the 
average $(B-V)$ =0.28 of the Hewitt \& Burbidge (1993) quasars with 
$0.7 \leq z \leq 3.0$ shows that, on average, $B = B_J + 0.07$ and $V = B_J - 0.21$. 
\\
\end{minipage}
\end{center}
\end{table}

\begin{table}[ht]
\small
\begin{center}
\caption{High- and Low-velocity Absorption Frequency in Narrow and Broad Line Quasars$^a$ 
\label{fwfreq1.tab}}
\begin{tabular}{lcccccc}
\\
\tableline
\tableline \\[-8pt]
{}&
\multicolumn{3}{c}{\underline{~~~~$|$Absorber Vel.$|$ $\leq$ 5000 \kms~~~~}}&
\multicolumn{3}{c}{\underline{~~~~$|$Absorber Vel.$|$ $>$ 5000 \kms~~~~}}\\
{}& {}& {N (FW$^b$ $\geq$}& {N (FW $<$}&
{}& {N (FW $\geq$ }& {N (FW $<$ }\\
{Sample}& {N}& {6000 km s$^{-1}$)}& {6000 km s$^{-1}$)}&
{N}& {6000 km s$^{-1}$)}& {6000 km s$^{-1}$)}\\
\multicolumn{1}{c}{(1)}  & {(2)} & {(3)} & {(4)} & {(5)} & {(6)} & {(7)} \\
\tableline \\[-10pt]
&\multicolumn{6}{c}{Number of absorbed quasars and counting errors}\\[3pt]
RQQ          & 12$\pm$3.5 & 6$\pm$2.4; 50$\pm$20\% & ~6$\pm$2.4; ~50$\pm$20\% & 10$\pm$3.2  & 2$\pm$1.4; 20$\pm$14\% & 8$\pm$2.8; 80$\pm$28\% \\
RLQ          & 19$\pm$4.4 & 5$\pm$2.2; 26$\pm$12\% & 14$\pm$3.7; ~74$\pm$19\% & 11$\pm$3.3  & 3$\pm$1.7; 27$\pm$15\% & 8$\pm$2.8; 73$\pm$25\% \\
CDQ          & ~4$\pm$2.0 & 0$\pm$0.0; ~0$\pm$~0\% & ~4$\pm$2.0; 100$\pm$50\% & ~3$\pm$1.7  & 1$\pm$1.0; 33$\pm$33\% & 2$\pm$1.4; 66$\pm$46\% \\
LDQ          & 12$\pm$3.5 & 4$\pm$2.0; 33$\pm$17\% & ~8$\pm$2.8; ~66$\pm$24\% & ~5$\pm$2.2  & 1$\pm$1.0; 20$\pm$20\% & 4$\pm$2.0; 80$\pm$40\% \\
CSS          & ~3$\pm$1.7 & 1$\pm$1.0; 33$\pm$33\% & ~2$\pm$1.4; ~66$\pm$46\% & ~3$\pm$1.7  & 1$\pm$1.0; 33$\pm$33\% & 2$\pm$1.4; 66$\pm$46\% \\
CSS$+$GPS    & ~3$\pm$1.7 & 1$\pm$1.0; 33$\pm$33\% & ~2$\pm$1.4; ~66$\pm$46\% & ~3$\pm$1.7  & 1$\pm$1.0; 33$\pm$33\% & 2$\pm$1.4; 66$\pm$46\% \\[3pt] 
\tableline \\[-8pt]
&\multicolumn{6}{c}{Number of absorbers and counting errors}\\[3pt]
RQQ          & 17$\pm$4.1 & 9$\pm$3.3; 53$\pm$19\% & ~8$\pm$2.8; ~47$\pm$17\% & 13$\pm$3.6  & 2$\pm$1.4; 15$\pm$11\% &11$\pm$3.3; 85$\pm$25\% \\
RLQ          & 23$\pm$4.8 & 8$\pm$2.8; 35$\pm$12\% & 15$\pm$3.9; ~65$\pm$17\% & 12$\pm$3.5  & 3$\pm$1.7; 25$\pm$14\% &~9$\pm$3.0; 75$\pm$25\% \\
CDQ          & ~4$\pm$2.0 & 0$\pm$0.0; ~0$\pm$~0\% & ~4$\pm$2.0; 100$\pm$50\% & ~3$\pm$1.7  & 1$\pm$1.0; 33$\pm$33\% &~2$\pm$1.4; 66$\pm$47\% \\
LDQ          & 14$\pm$3.7 & 6$\pm$2.4; 43$\pm$17\% & ~8$\pm$2.8; ~57$\pm$20\% & ~6$\pm$2.4  & 1$\pm$1.0; 17$\pm$17\% &~5$\pm$2.2; 83$\pm$37\% \\
CSS          & ~5$\pm$2.2 & 2$\pm$1.4; 40$\pm$28\% & ~3$\pm$1.7; ~60$\pm$34\% & ~3$\pm$1.7  & 1$\pm$1.0; 33$\pm$33\% &~2$\pm$1.4; 66$\pm$47\% \\
CSS$+$GPS    & ~5$\pm$2.2 & 2$\pm$1.4; 40$\pm$28\% & ~3$\pm$1.7; ~60$\pm$34\% & ~3$\pm$1.7  & 1$\pm$1.0; 33$\pm$33\% &~2$\pm$1.4; 66$\pm$47\% \\[3pt] 

\tableline
\tableline
\end{tabular}
\end{center}
\begin{center}
\begin{minipage}{16cm}
\footnotetext{ }{$ a$ Note: The statistics in this table are based only on absorbers with restframe 
$EW \geq$ 0.5\AA{} and \mbox{$|$ abs. vel. $| \leq$21\,000\,\kms{}}, the most complete subset 
of absorbers (see \S~\ref{compltness}). Note that quasars with both high and low velocity
absorbers will be counted both in columns 2 and 5. Hence, the sum of the entries in these columns
may exceed the number of absorbed quasars listed in Tables~\ref{freq.tab} (top section, column 3 right) 
and~\ref{fwfreq2.tab} (columns 6 and 7).}\\
\footnotetext{ }{$ b$ FWHM of the \civ{} {\em emission} line.} \\
\end{minipage}
\end{center}
\end{table}

\vspace{0.50cm}
\begin{table}[h]
\small
\begin{center}
\hspace{0.30cm}
\caption{Frequency of Absorbed and Non-Absorbed Narrow and Broad Line Quasars$^a$
\label{fwfreq2.tab}}
\begin{tabular}{lcccccc}
\\
\tableline
 \tableline\\[-8pt]
{  }& {\# } & {\# }& \\
{  }& {QSOs} & {QSOs}&
\multicolumn{2}{c}{\underline{~~~~~~~~~Non-Absorbed Quasars~~~~~~~~~} }&
\multicolumn{2}{c}{\underline{~~~~~~~~~~~~~~~Absorbed Quasars~~~~~~~~~~~}}\\
{Sample} & (NL$^b)$ & (BL$^c$) & {\# QSOs (NL)} &{\# QSOs (BL)} &{\# QSOs (NL)} &{\# QSOs (BL)}\\
\multicolumn{1}{c}{(1)}  & (2) & (3) & (4) & (5) & (6) & (7) \\
\tableline \\[-8pt]
All QSOs     & 91 & 23 & 58$\pm$7.6; 64$\pm$~8\% & 11$\pm$3.3; 48$\pm$15\% & 33$\pm$5.7; 36$\pm$\,~6\% & 12$\pm$3.5; ~52$\pm$~15\% \\
RQQ          & 34 & 14 & 21$\pm$4.6; 62$\pm$13\% & ~8$\pm$2.8; 57$\pm$20\% & 13$\pm$3.6; 38$\pm$10\%   & ~6$\pm$2.4; ~43$\pm$~17\% \\
RLQ          & 57 & ~9 & 37$\pm$6.1; 65$\pm$11\% & ~3$\pm$1.7; 33$\pm$19\% & 20$\pm$4.5; 35$\pm$\,~8\% & ~6$\pm$2.4; ~66$\pm$~26\% \\
CDQ          & 15 & ~3 & 10$\pm$3.2; 67$\pm$21\% & ~2$\pm$1.4; 66$\pm$46\% & ~5$\pm$2.2; 33$\pm$15\%   & ~1$\pm$1.0; ~33$\pm$~33\% \\
LDQ          & 27 & ~4 & 16$\pm$4.0; 59$\pm$15\% & ~0$\pm$0.0; ~0$\pm$\,~0\% & 11$\pm$3.3; 41$\pm$12\%   & ~4$\pm$2.0; 100$\pm$~50\% \\
CSS          & 13 & ~1 & ~9$\pm$3.0; 69$\pm$23\% & ~0$\pm$0.0; ~0$\pm$\,~0\% & ~4$\pm$2.0; 31$\pm$15\%   & ~1$\pm$1.0; 100$\pm$100\% \\
CSS$+$GPS    & 16 & ~2 & 11$\pm$3.3; 69$\pm$20\% & ~1$\pm$1.0; 50$\pm$50\% & ~4$\pm$2.0; 27$\pm$13\%   & ~1$\pm$1.0; 100$\pm$~50\%\\[3pt] 
\tableline
\tableline
\end{tabular}
\end{center}
\begin{center}
\begin{minipage}{16cm}
\footnotetext{}{$ a$ Here a quasar is considered to be absorbed if the
individual \civ{} absorption doublets has restframe $EW \geq 0.5$ \AA{} 
and its absolute velocity is less than 21\,000\,\kms{} (the
most complete subset of absorbers; see text). As a result those quasars 
{\it with} \civ{} absorption which does not satisfy these conditions are 
here counted as non-absorbed.} \\
\footnotetext{ }{$ b$ NL: Quasars with narrow {\em emission} lines: FWHM(\civ{}) $<$ 6000 \kms.}\\
\footnotetext{ }{$ c$ BL: Quasars with broad {\em emission} lines: FWHM(\civ{}) $\geq$6000 \kms.}
\end{minipage}
\end{center}
\end{table}

\end{document}